\documentclass[useAMS,usenatbib]{mn2e}

\usepackage{graphicx}
\usepackage{amssymb}
\usepackage{amstext}
\usepackage{url}
\usepackage{amsmath}
\usepackage{setspace}
\usepackage{xtab}
\usepackage{textcomp}
\usepackage{booktabs,array}
\usepackage{color}
\usepackage{colortbl}
\usepackage{tabularx}
\usepackage{gensymb}
\usepackage{subfigure}

\title[Mode identification of $\gamma$ Doradus star HD\,135825]{Spectroscopic Pulsational Frequency
Identification and Mode Determination of $\gamma$ Doradus Star HD\,135825}
\author[E. Brunsden, K. R. Pollard, P. L. Cottrell, D. J. Wright, P. De Cat, P.
M. Kilmartin]{E. Brunsden$^{1}$\thanks{E-mail: emily.brunsden@gmail.com}, K. R.
Pollard$^{1}$, P. L. Cottrell$^{1}$, D. J. Wright$^{2}$\thanks{E-mail:
duncan.wright@unsw.edu.au}, P. De Cat$^{3}$ \and P. M. Kilmartin$^{1}$ \\
$^{1}$Department of Physics and Astronomy, University of Canterbury, Private Bag
4800, Christchurch, New Zealand\\
$^{2}$Department of Astrophysics, University of New South Wales, Sydney,
Australia\\
$^{3}$Royal Observatory of Belgium, Ringlaan 3, 1180 Brussel, Belgium}
\begin{document}

\date{ }

\pagerange{\pageref{firstpage}--\pageref{lastpage}} \pubyear{2012}

\maketitle

\label{firstpage}

\begin{abstract}
We present
the mode identification of frequencies found in spectroscopic observations
of the $\gamma$ Doradus star HD\,135825. Four frequencies were successfully
identified: $1.3150$~$\pm$~$0.0003$~d$^{-1}$; $0.2902$~$\pm$~$0.0004$~d$^{-1}$; $1.4045$~$\pm$~$0.0005$~d$^{-1}$; and
$1.8829$~$\pm$~$0.0005$~d$^{-1}$. These correspond to ($l$, $m$) modes of (1,1), (2,-2), (4,0)
and (1,1) respectively. Additional frequencies were found but they were below the signal-to-noise limit of the Fourier spectrum and not
suitable for mode identification. The rotational axis
inclination and $v$sin$i$ of the star were determined to be $87 \degree$ (nearly edge-on) and $39.7$
kms$^{-1}$ (moderate for $\gamma$ Doradus stars) respectively.
A simultaneous fit of these four modes to the line profile variations in the data gives a reduced $\chi^2$ of $12.7$. We confirm, based on the
frequencies found, that HD\,135825 is a bona fide
$\gamma$ Doradus star. 
\end{abstract}

\section{Introduction}
\thispagestyle{empty}
The pulsation of a star is reflective of its internal structure. Since the
interior of a star cannot be directly seen, the analysis of the pulsation is the
most advanced way to determine interior properties and dynamics. Because pulsations
occur in many different types of stars, in many stages of stellar lifetimes,
asteroseismology can be used to further our understanding of stellar evolution.

Asteroseismology is the study of vibrational physics in stars, specifically pulsational behaviour in order to derive interior structure
properties. The characteristics of pulsation
are described by the ``quantum numbers'' $n$, $l$ and $m$. The radial order,
$n$, describes the number of interior nodal shells, $l$ gives the
number of nodal lines on the stellar surface and the azimuthal number, $m$,
defines the number of nodal lines that intersect the pole.  

By analysing the non-radial pulsations and determining the spherical harmonic pulsation modes, we obtain asteroseismic information
about the deep interior regions of stars. Non-radial pulsations are classified by the restoring force of the motion: gravity (g-mode) or
pressure (p-mode). Gravity-mode pulsations propagate as deep as the convective core interface making them ideal probes of core physics. The detailed
frequency spectrum of excited modes places severe constraints on the physical conditions within the regions where these modes propagate. Successful
mode identifications can be found in \citet{2004MNRAS.347..463A}, \citet{2006A&A...455..235Z} and
\citet{2009A&A...497..183B} for various types of pulsating star.

The study presented here focuses on one g-mode pulsator, the $\gamma$ Doradus
candidate star HD\,135825. The defining feature of the $\gamma$
Dor class is the presence of high-order non-radial g-mode pulsations in an A-F type star that is on or near the main sequence (see
\citealt{1999PASP..111..840K} and \citealt{2007CoAst.150...91K}, \citealt{2009AIPC.1170..455P} for reviews). Thus these stars are slightly hotter and
slightly larger than the Sun. A typical feature of the $\gamma$ Dor class is the presence of multiple g-mode pulsations at frequencies lower than
the fundamental radial mode \footnote{Typical radial pulsation frequencies for A-F stars are 8-24 d$^{-1}$}. Observed g-mode frequencies generally
fall between 0.3 - 3.3 d$^{-1}$, although the observed frequencies are dependent on
the stellar rotation and direction of the travelling wave.  The pulsation frequency range for $\gamma$ Dor stars is similar to the  g-mode pulsators
of
the Slowly Pulsating B (SPB) class, so an estimate of the T$_{eff}$ or spectral class is required in order to distinguish between these classes.
Presently there are less than 100 bright bona fide $\gamma$ Dor stars (see list in \citealt{2011AJ....142...39H}) with a further 100 candidates thus
far reported by Kepler \citep{2010ApJ...713L.192G,2011A&A...534A.125U}.

The $\gamma$ Dor pulsators reside in that part of the Hertzsprung-Russell diagram
between the classical instability strip and the solar instability region. Thus they border the regions of the $\delta$ Scuti (p-mode) pulsators and
the solar-like oscillators. A few hybrid $\gamma$ Dor/$\delta$ Scuti stars have now been studied (e.g. \citealt{2005AJ....129.2026H},
\citealt{2008AandA...489.1213U}) and further candidates proposed and still being discovered by Kepler
\citep{2011A&A...534A.125U,2011MNRAS.417..591B}. Early indications show that $\gamma$ Dor/$\delta$ Scuti stars may challenge our understanding of the
instability strip \citep{2011A&A...534A.125U}. 

Solar-like variability is driven in the convective
envelope of a star from stochastic excitation and has short frequencies (e.g. the F-type star Procyon has solar-like variability frequencies in the
range 300-1400 $\micro$Hz, equivalent to 26-121 d$^{-1}$,  \citealt{2010ApJ...713..935B}). Solar-like oscillations also exist in stars slightly
hotter than the Sun \citep{2008Sci...322..558M} and
have been discovered in the $\delta$ Scuti star HD\,187547 \citep{2011Natur.477..570A}. This indicates that the convective envelope operates
efficiently in stars up to 2M$_\odot$. It is possible that in $\gamma$ Dor stars, being located between solar-like stars and $\delta$ Scuti stars in
size and temperature, solar-like oscillations may also be present. No solar-like/$\gamma$ Dor hybrids have yet been discovered but it may be possibile
for high-resolution satellite photometry to be able to detect the high frequency pulsations.

The multitude of high-order, low-degree g-modes present in $\gamma$ Dor stars give us information about their interior structure and
physical conditions. These stars have both a convective envelope and a convective core, in between which a radiative zone exists. Driving occurs by
the ``flux-blocking mechanism'' where convection at the base of the thin convective envelope modulates flux throughput from the core to the stellar
atmosphere \citep{2000ApJ...542L..57G, 2004A&A...414L..17D}. Asteroseismic study of interior structure requires frequencies to be identified and
corresponding modes to be fully characterised with ($l$,$m$) values. Mode analysis
can also be used to constrain stellar parameters such as inclination \citep{2011ApJ...728L..20W}.

High precision photometry has
been successful in identifying up to hundreds of frequencies in $\gamma$ Dor stars (see \protect\cite{2011A&A...525A..23C} for an
example of 840 identified frequencies using the COROT photometric satellite). Photometry is invaluable in the study of $\gamma$ Dor
pulsations as it can be used to find such large numbers of frequencies with high precision and also determine the corresponding $l$ values. 
Spectroscopy has thus far been limited to ground-based observations and the signal-to-noise requirements for study of spectra have meant longer
integration times. The benefit of spectroscopic observations is the determination of both $l$ and $m$ values for each pulsational frequency.  As such
photometric and spectroscopic studies of these stars are complementary. To date only a handful of $\gamma$ Dor stars have spectroscopic mode
identifications published with only a few (less than six) frequencies each. The lack of detailed mode identifications has previously limited interior
structure studies but it is
hoped that identifications such as in this paper can be incorporated and used to refine current stellar models.

A further aspect of interest in $\gamma$ Dor stars is the effect of rotation on pulsation. Current models assume slow rotation with no effect on
the pulsational geometry. The Coriolis and centrifugal accelerations become more important with increased rotation. Rotational effects can lead
to differential rotation and, in extreme cases, the breakdown of spherical symmetry
\citep{2003MNRAS.343..125T,2006A&A...455..607L,2006A&A...455..621R}. It is still not well
understood to what extent rotation affects spectroscopic observations. Rotation can further hinder frequency analysis as $\gamma$ Dor
stars
have rotational periods of the same order as their pulsational periods.

This study focuses on one member, HD\,135825, of the $\gamma$ Dor group of stars to identify the
frequencies and modes of the g-mode pulsations present and to look for any observational evidence of hybrid pulsation and rotational
effects. Section \ref{obs} discusses the specific data acquisition and reduction undertaken. The analysis of the spectra follows in Sections
\ref{frequency} and \ref{modeid}. Specifically, Section \ref{frequency} breaks down each of the frequency analysis processes.
Section \ref{modeid} then discusses the corresponding mode identifications to the frequencies detected. A discussion of results in context follows in
Section \ref{disc}.

\section{Observations and data treatment}\label{obs}

In total nearly 300 spectra of HD\,135825 were collected from the 1-metre
McLellan telescope at Mt John University Observatory (MJUO) in Tekapo, New
Zealand
($170\degree 27'.9$~E, $43\degree 59'.2$~S). The observatory is at an elevation
of $1029$~m. Spectra were obtained on the fibre-fed High Efficiency and
Resolution Canterbury University Large Echelle Spectrograph (HERCULES). HERCULES
operates over a wavelength range of 3800-8000~\AA\ \citep{2003ASPC..289...11H} and
the $4096$x$4096$ pixel CCD installed in 2007 samples the entire free spectral
range in one exposure. 

Multi-site data has also been taken for this star. A total of 28 spectra were taken at on the Sadiford Cass Echelle Spectrometer on the 2.1 m
telescope
at McDonald Observatory, USA. In addition 8 spectra were obtained on the SOPHIE spectrograph on the 193cm Telescope at the Observatoire de
Haute Provence, and 3 spectra  on the McKellar Spectrograph on the 1.2-metre telescope at the Dominion Astrophysical Observatory. In order to
incorporate these data a mean line profile with sufficient signal-to-noise for pulsation analysis must be constructed. The small size of the
aforementioned datasets did not allow for a reliable mean profile with sufficient phase coverage to be produced. These data were therefore not
incorporated in the present analysis. 

Data from MJUO were collected over a period of $18$ months from February 2009 to July
2010. These data were reduced using a MATLAB pipeline written by Dr. Duncan
Wright. The pipeline performs the basic steps of flat fielding from white-lamp
observations, calculating a dispersion solution from thorium-lamp observations
and outputting the data into a two-dimensional format. The observations were
then processed in a pipeline written specifically for non-radial pulsation
analysis. The processing stage corrects for barycentric motion, removes
small differences between observations, fits the continuum using a synthetic spectrum, then normalises and merges the orders
for each spectroscopic observation.

Each image was then cross-correlated using the delta-function method (Wright et
al. (in preparation), building on \cite{NewEntry3}, \cite{2007CoAst.150..135W}) to
create a representative line-profile. The delta-function method relies on the following
assumptions of the spectra: spectral lines at different wavelengths have the
same shape, lines of different species and different excitation potentials
change in the same way as a result of the pulsation and all lines in a spectrum are
varying in phase. The first two criteria were shown to hold to a
good approximation in \cite{NewEntry3} and a check of lines of
various equivalent width, species, excitation potential, wavelength and  showed them to be varying consistently
in phase in a single spectrum for HD\,135825. 

Time-series analysis and
mode identification was done using the FAMIAS
software \citep{2008CoAst.157..387Z}, applied as in \citet{2006A&A...455..235Z}. This entailed an analysis of
the moments 
\citep{1986MNRAS.219..111B,2003AandA...398..687B} and pixel-by-pixel
variations which identified the frequencies in the spectral lines. Specifically, FAMIAS uses the moment definition described in
\citet{1992AandA...266..294A} and revised in \citet{2003AandA...398..687B}.

The software package SigSpec \citep{2007AandA...467.1353R} was used as a validation of the frequency selection
method. SigSpec performs a Fourier analysis of a two-dimensional dataset
and selects frequencies based on their spectral
significance.

Mode identification was done using the Fourier Parameter Fit method. Developed
by Wolfgang Zima \citep{2009AandA...497..827Z}, this extension of the
pixel-by-pixel frequency analysis method uses each identified frequency in the
line profile variation and matches this to synthetic spectra with input stellar
parameters for various modes until the best fit is obtained.

\section{Frequency Analysis}\label{frequency}

HD\,135825 is a relatively faint ($V=7.3$), and hence comparatively understudied,
$\gamma$ Dor candidate
star. It has a moderate $v$sin$i$ of $38 \pm 5$\,kms$^{-1}$
\citep{2006AandA...449..281D}. The star was confirmed multi-periodic in
photometry by \citet{2002ASPC..256..203E} and has an effective temperature of
$7050 \pm 90$\,K and log $g$ of $4.39 \pm 0.13$, both from \citep{2008AandA...478..487B} and spectral type
F0 \citep{1999IBVS.4659....1K}. A full abundance analysis has been done using high-resolution, broad wavelength range spectra by
\citet{2008AandA...478..487B} with a Metallicity $[M/H]= +0.13 \pm 0.09$ found. Full abundance results are available in the on-line material. One
pulsation period has
previously been identified in
photometry, $0.76053$~d ($1.31487$~d$^{-1}$), from the HIPPARCOS satellite data
and one best fitting frequency from classification spectroscopy, $0.63$~d ($1.59$~d$^{-1}$), both by
\citet{2006AandA...449..281D}. Note the main period from photometry was also suggested as the main period in \citet{2006AandA...449..281D}
spectroscopy but was not recoverable due
to the limited size of the radial velocity dataset.

Each of the frequency variation tests (moments 0-3 and pixel-by-pixel) was
used to construct periodograms for the star. Some methods have higher
signal-to-noise and thus improve our ability to extract more frequencies. Amplitudes tabulated in this paper are scaled to ratios of $f_1$ to allow
comparison between identification methods with differing absolute amplitudes.

\begin{center}
\begin{figure}
   \includegraphics[width=0.5\textwidth]{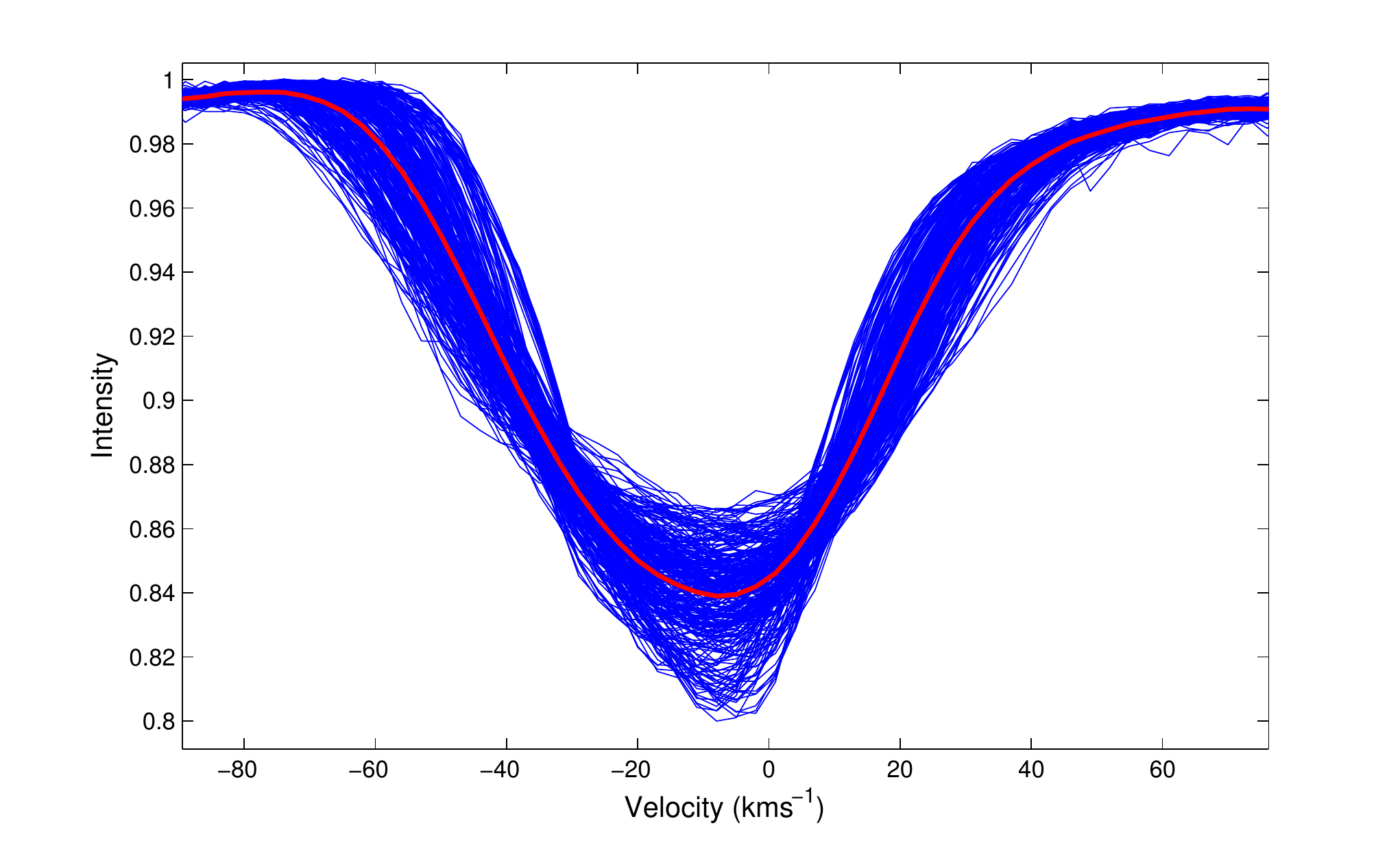}
\caption{Zero-point profile for the 291 observations of HD\,135825 with the mean
shown in red. The smooth mean shows that the variations in the profile are well sampled. The line profile data are provided as an on-line file that
includes the Modified Julian Date, velocity (on a relative scale) and the intensity at each of the 800 velocity sampling points for each profile.
}\label{zpprofile}
\end{figure}
\end{center}

Representative line profiles  are created by cross-correlating
hundreds of spectral lines. These profiles of HD\,135825 are shown in
Figure \ref{zpprofile} and are analysed using the
pixel-by-pixel and zeroth to third moments.
These tests cover all the independent frequencies. In addition to these
tests,
the software SigSpec was used to analyse all the two-dimensional data sets (all
the moments) and also each individual pixel across the line profile, to confirm
the frequencies found and their significance.

The window function of the dataset was calculated to show any periodicities
intrinsic to the data sampling. The window function for HD\,135825 is plotted
in Figure \ref{window}. As the dataset is from a single site the one-day
aliasing is obvious and this can present a problem in the identification of
frequencies. In this analysis, care has been taken to consider all one-day
aliases and determine the true frequency. 

\subsection{Pixel-by-Pixel}

\begin{figure}
\subfigure[]{
\includegraphics[width=0.5\textwidth]{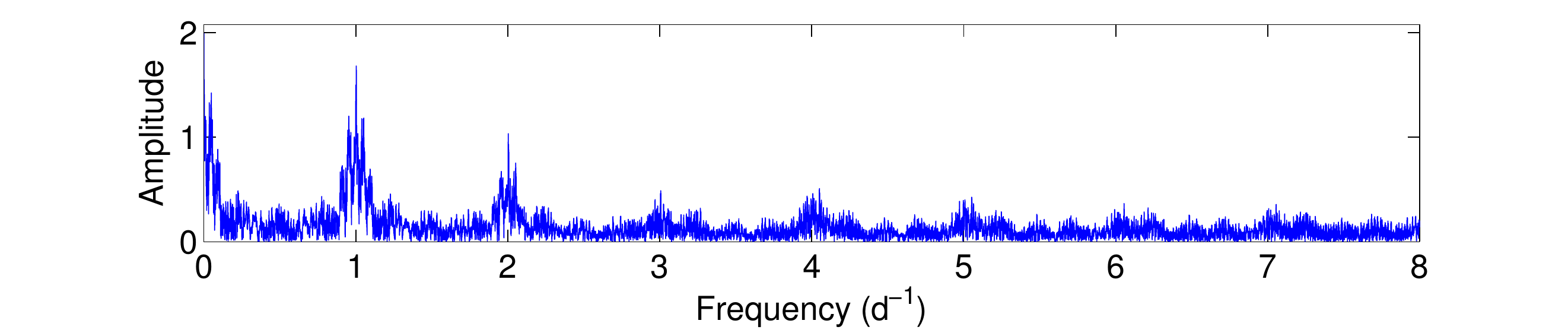}\label{window}
}
   \subfigure[]{
\includegraphics[width=0.5\textwidth]{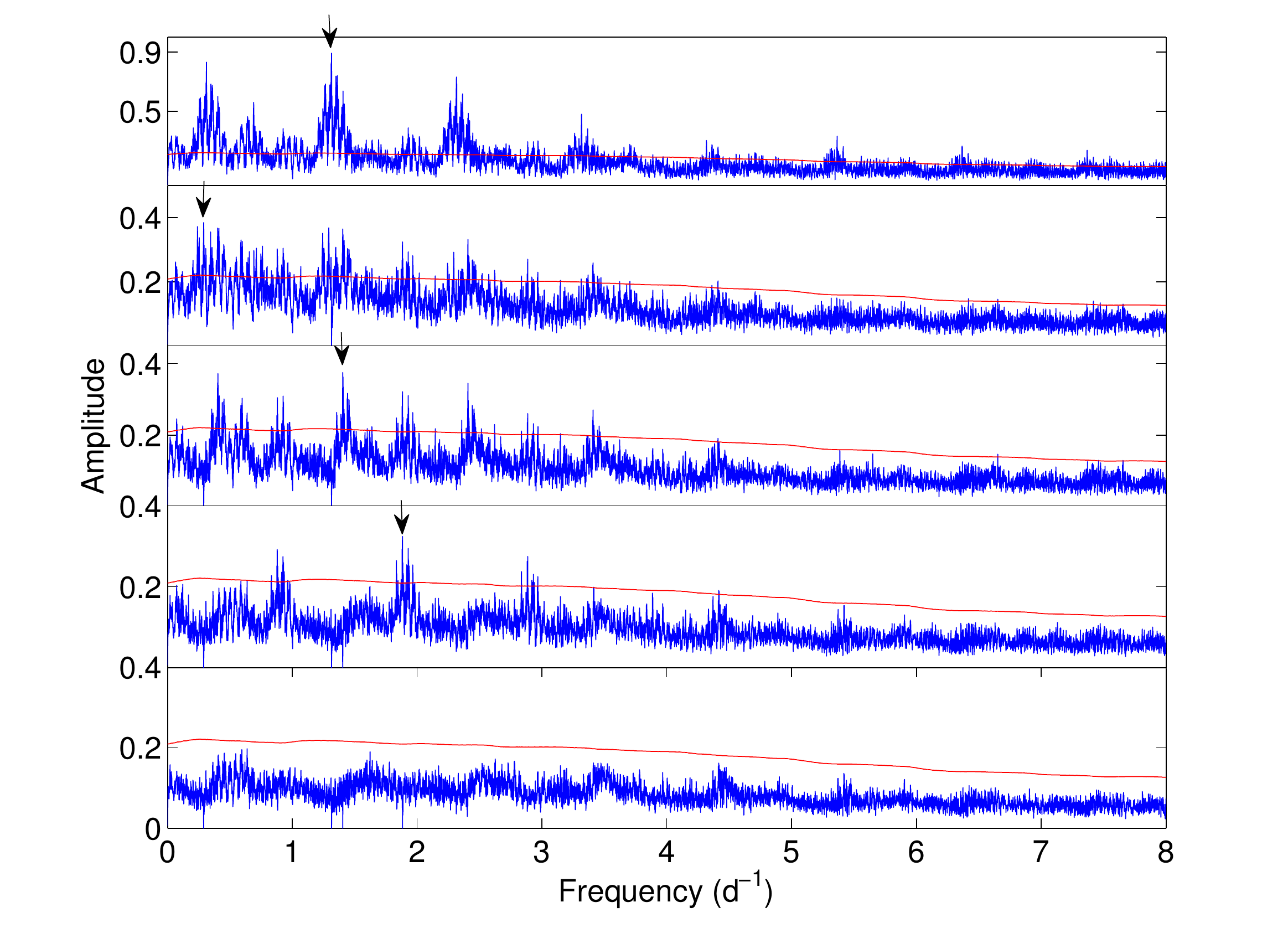}\label{pbpfreq}
}
\caption{The
Fourier spectra showing the amplitude of frequencies detected in the data. \subref{window} Window function for the 291 observations of
HD\,135825. \subref{pbpfreq} frequencies found using the pixel-by-pixel
technique and successive pre-whitening. The smooth line shows the significance
level taken to be the limit of detection.}
\end{figure}

\begin{figure}
\subfigure[The dataset from MJUO phased to $f_{p1}=1.3149$ d$^\protect{-1\protect}$]{
\includegraphics[scale=0.45]{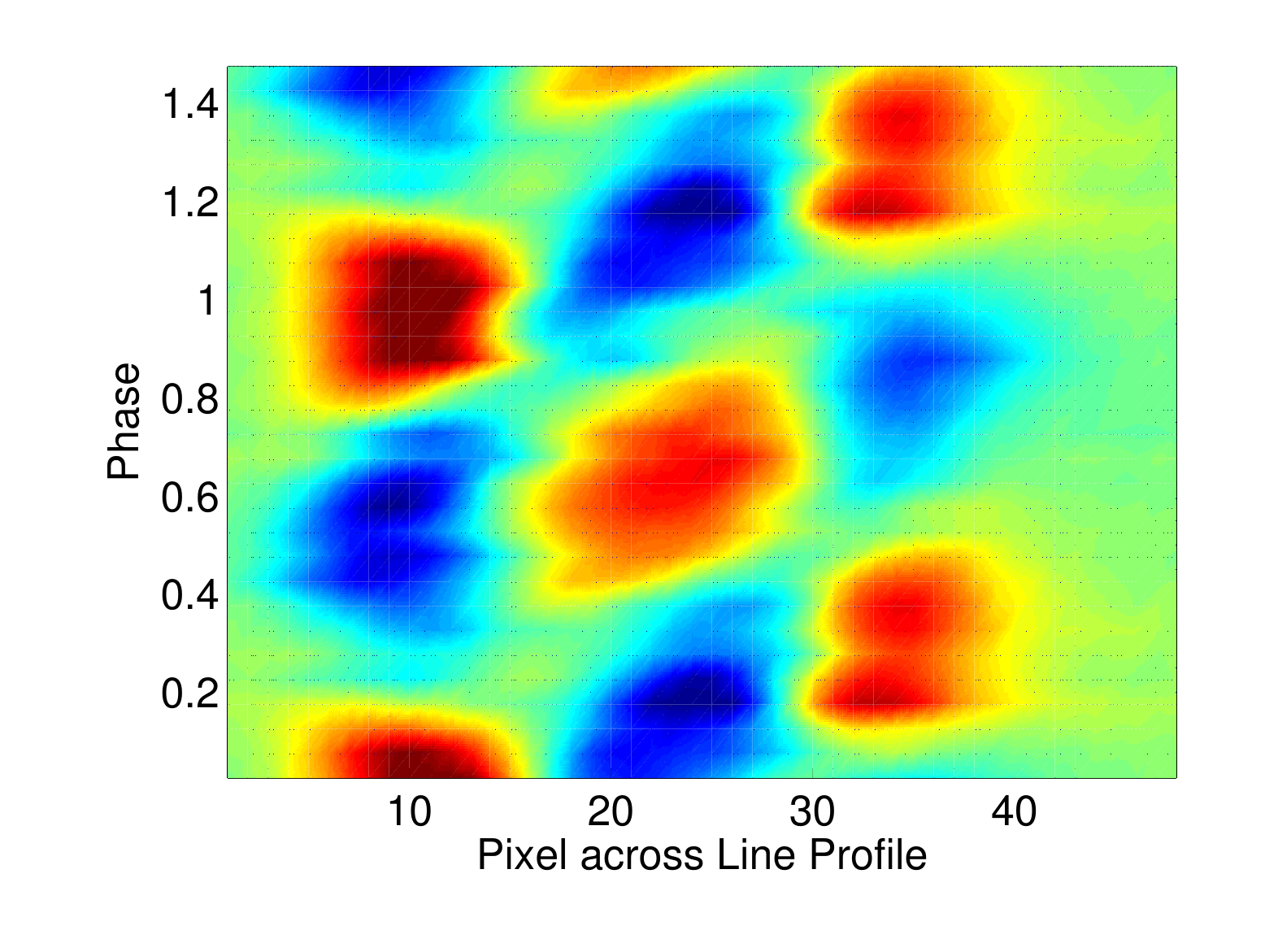}
\label{phadata}
}
 \subfigure[Fit with $f_{p1}$ $(1.3149 $d$^\protect{-1\protect})$]{
\includegraphics[scale=0.25]{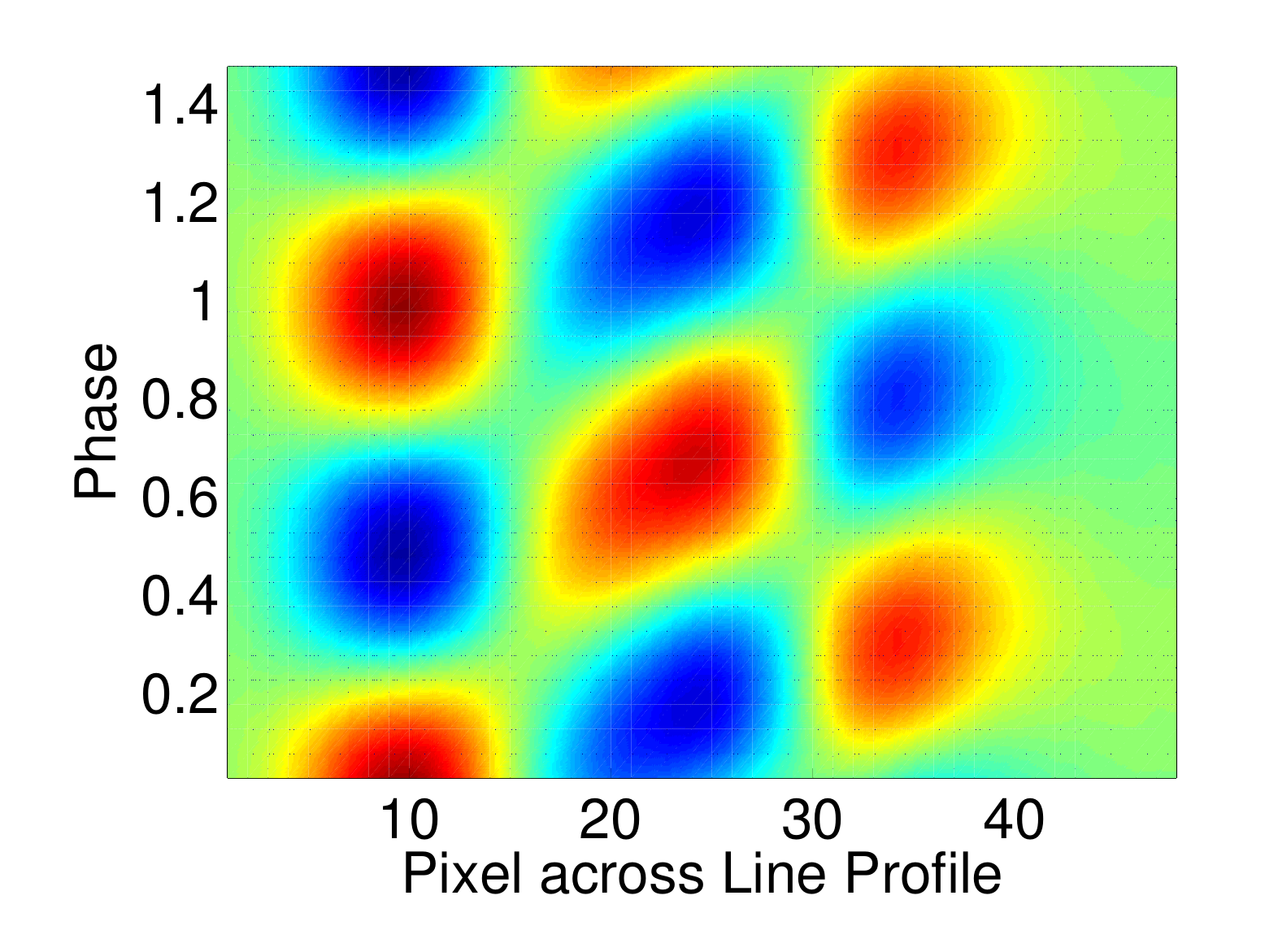}
\label{phafit1}
}
\subfigure[Fit with $f_{p2}$ $(0.2901 $d$^\protect{-1\protect})$]{
\includegraphics[scale=0.25]{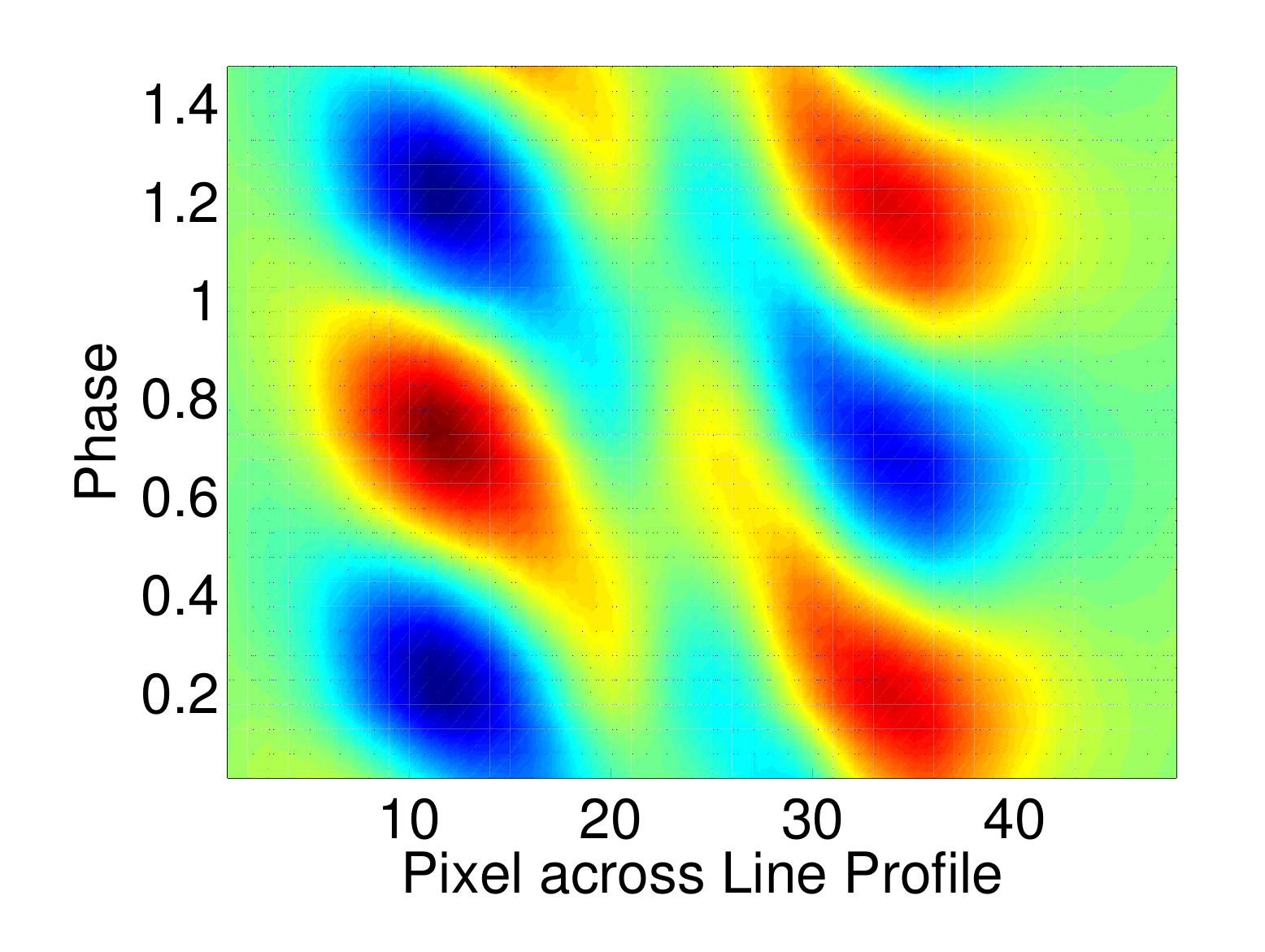}
\label{phafit2}
}
\subfigure[Fit with $f_{p3}$ $(1.4045 $d$^\protect{-1\protect})$]{
\includegraphics[scale=0.25]{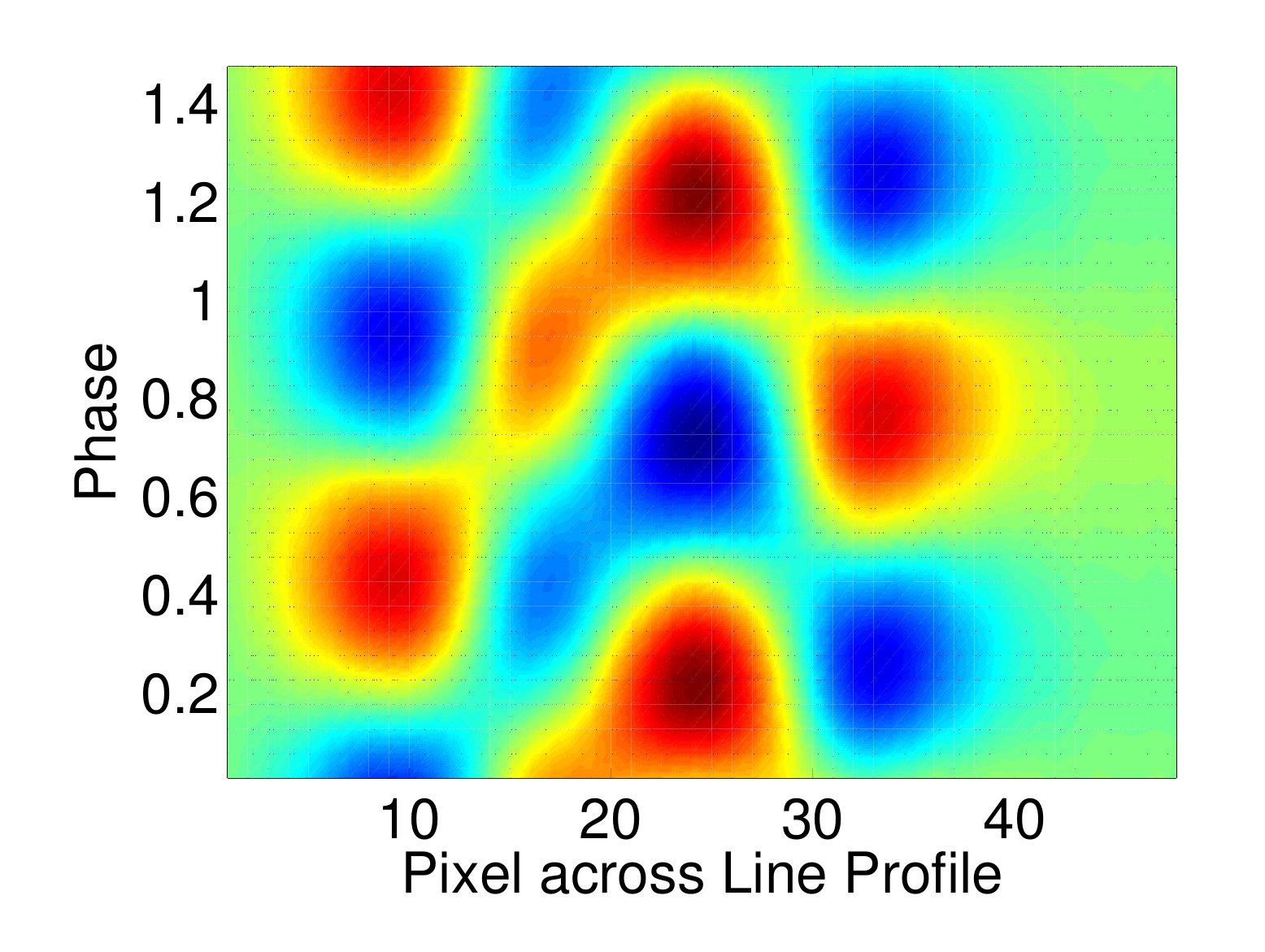}
\label{phafit3}
}
\subfigure[Fit with $f_{p4}$ $(1.8830 $d$^\protect{-1\protect})$]{
\includegraphics[scale=0.25]{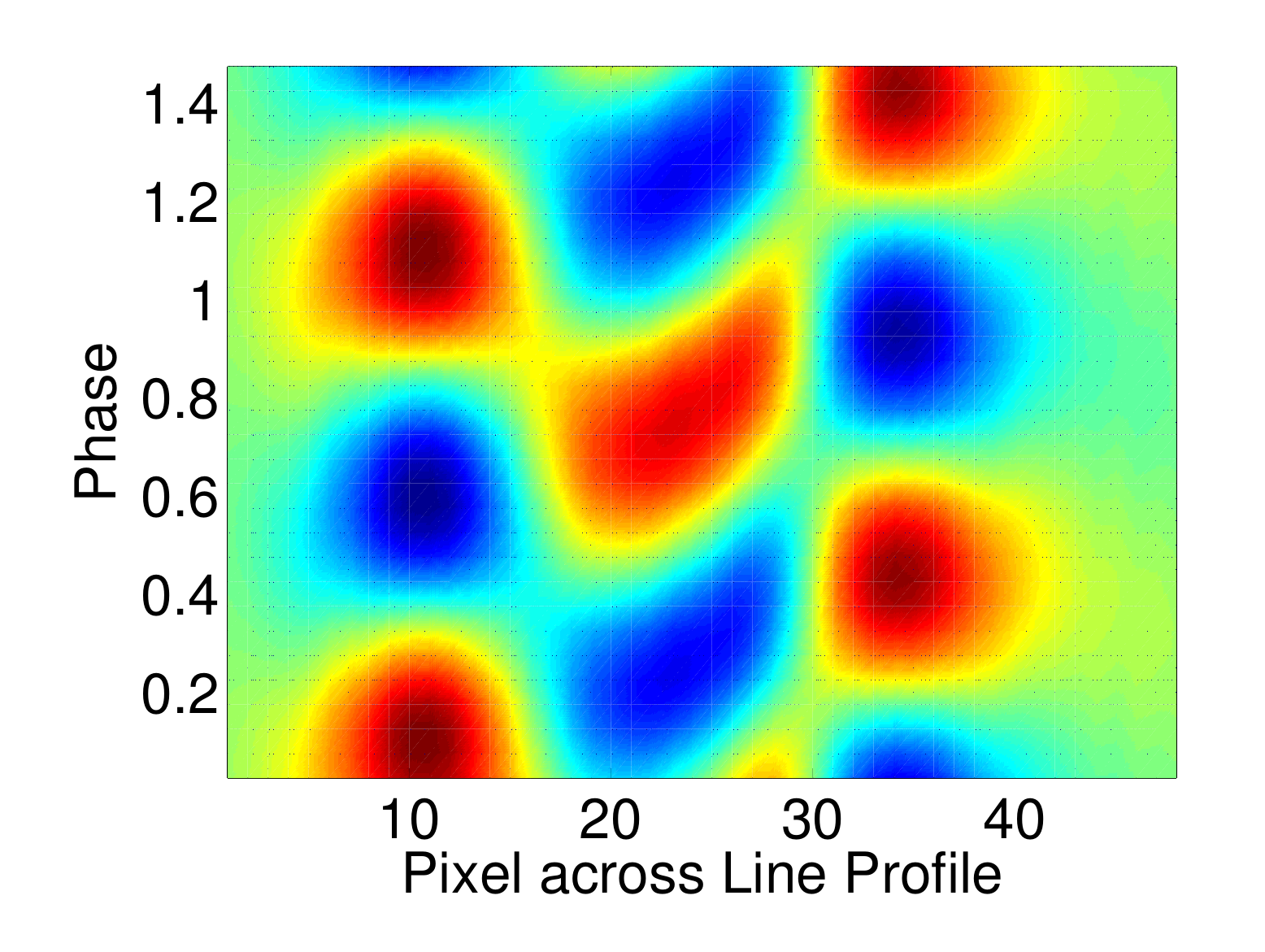}
\label{phafit4}
}
\label{phasedfits}
\caption[Residual phased line profiles and synthetic fits to the profiles using the frequencies and modes
identified]{Residual (line profile minus the mean line profile) phased line profiles and fits to the profiles using frequencies (in
d$^{-1}$) identified with the pixel-by-pixel method. \subref{phadata} Real dataset and \subref{phafit1} to
\subref{phafit4} are the fits to the profiles with the addition of successive
frequencies and modes constructed as described in Section \ref{synth}.}
\end{figure}

The pixel-by-pixel method is an effective tool for examining the frequencies and
their fits to the data in two-dimensions as this method performs an
analysis of the zero-point line profile by considering the variation of each pixel. The cross-correlated line profile of HD\,135825 contained 48
pixels.
The pixel-by-pixel technique produced the lowest noise-level in the Fourier
spectra. The Fourier spectra for the frequencies ($f_p$) are given in Figure
\ref{pbpfreq} and numerical results in Table \ref{pbpres}.
The phased variations of the zero-point profile are shown in Figure \ref{phadata}
and the fits for the four frequencies in Figures
\ref{phafit1}-\ref{phafit4}. The
profiles are smoothed and
phased to show the variation over $1.4$ cycles. Where
a bin has more than one observation for a particular phase a signal-weighted mean has
been used.
These plots show the increased matching to the data shape for each successive
additional
Frequency with the first frequency dominating the shape and the fit. 

Table \ref{pbpres} column five shows the amount of variation explained by each frequency
combination calculated from the residuals of the fits to the data at each pre-whitening stage. The percentage of the variation explained in
the
pixel-by-pixel method is generally lower due to several factors. Firstly we are considering the motion of all 48 pixels in the line profile, only
about 15 of which are strongly varying (see the amplitude variation profiles as in Figure \ref{modef1}). This means we are considering high and low
signal-to noise regions of the line profile. Another effect of considering all pixels in the line profile is the increased precision required to match
all the individual pixels. Frequencies, amplitudes and phases must be much more precise for an accurate sum of pixel fits to account for the movement
in the line profile. Despite this the Fourier spectra and the reduction of residuals in each pre-whitening stage leaves us confident in the
identification of these frequencies. With the fit of four frequencies, some 29$\%$ of the variability in the line profile remains.  Much of
the remaining variation is likely due to further un-identified frequencies at much lower amplitudes. The signal-to-noise limit of our data (see the
red
line in Figure \ref{pbpfreq}) restricts the extraction of further frequencies.

\begin{center}
\begin{table}\caption[Pixel-by-Pixel Frequencies]{Frequencies from the
pixel-by-pixel analysis of HD\,135825 from MJUO.}\label{pbpres}
\begin{center}
\begin{tabular}{c c c c c}
\\
\hline
ID & Frequency & Amplitude & Phase & Cumulative\\
& ($d^{-1}$) & (scaled to $f_{p1}$) & &Variation Explained \\
 \hline
 $f_{p1}$ & 1.3149 & 1 & 0.1033 & 49\% \\
 $f_{p2}$ & 0.2901 & 0.4786 & 0.4199 & 58\% \\
 $f_{p3}$ & 1.4045 & 0.4717 & 0.0368 & 66\%\\
 $f_{p4}$ & 1.8830 & 0.3874 & 0.4058 & 71\%\\ 
\hline
\end{tabular}
\end{center}
\end{table}
\end{center}

\subsection{Zeroth Moment (Equivalent Width)}

Variations in the equivalent width of the line profile are not expected
unless there are significant temperature variations, which makes the analysis much more complex. For HD\,135825 only one
frequency could be extracted ($1.3147$~d$^{-1}$), with a low-amplitude Fourier
peak above the noise base. The lack of frequencies in the zeroth moment indicates that there are 
only small temperature variations centred around the primary frequency and
that this star is a good candidate for spectroscopic pulsation analysis. 

\subsection{First Moment (Radial Velocity)}

\begin{figure}
\centering
\subfigure[]{
\includegraphics[width=0.4\textwidth]{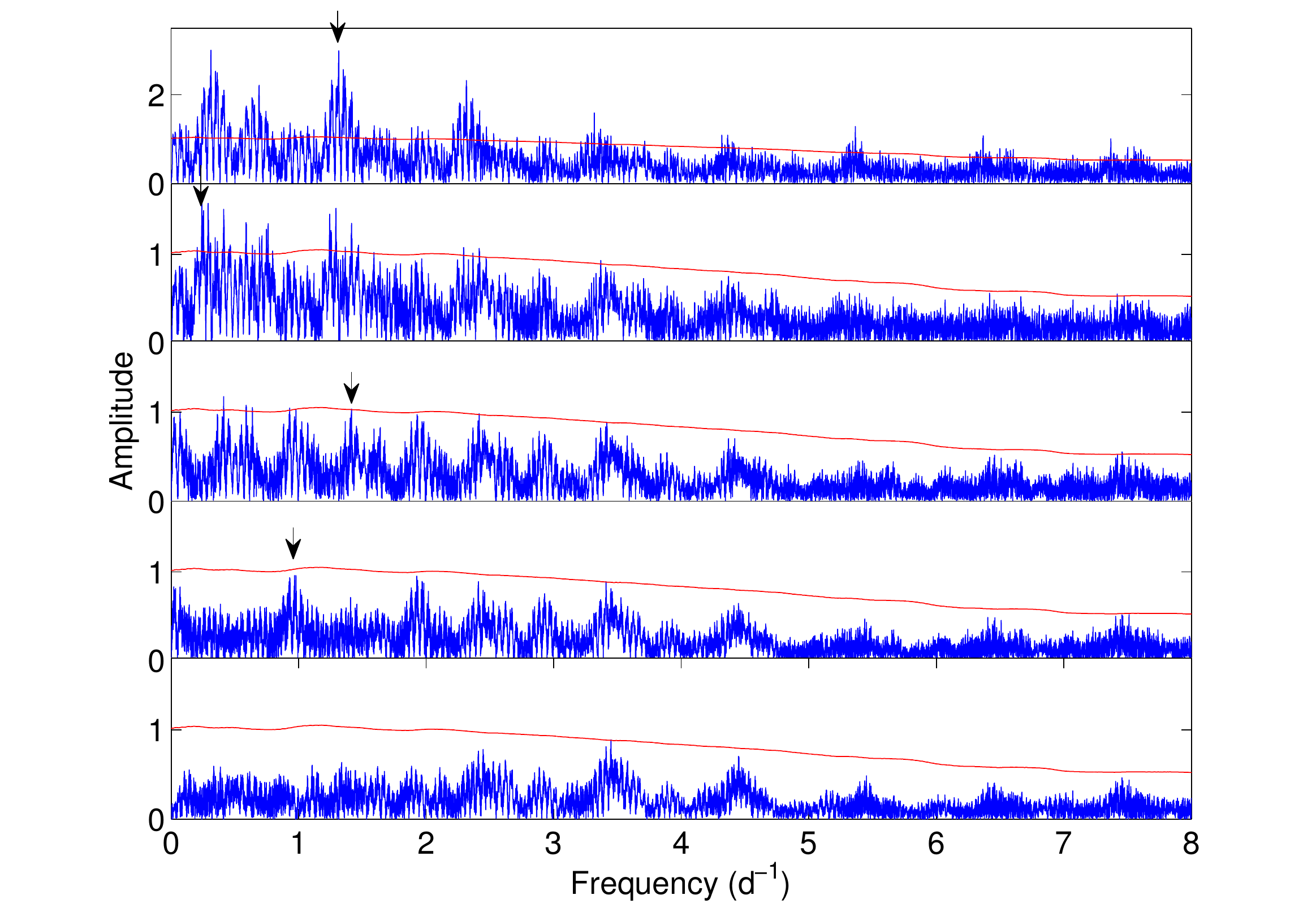}
\label{1momfreqa}
}
\subfigure[]{
\includegraphics[width=0.4\textwidth]{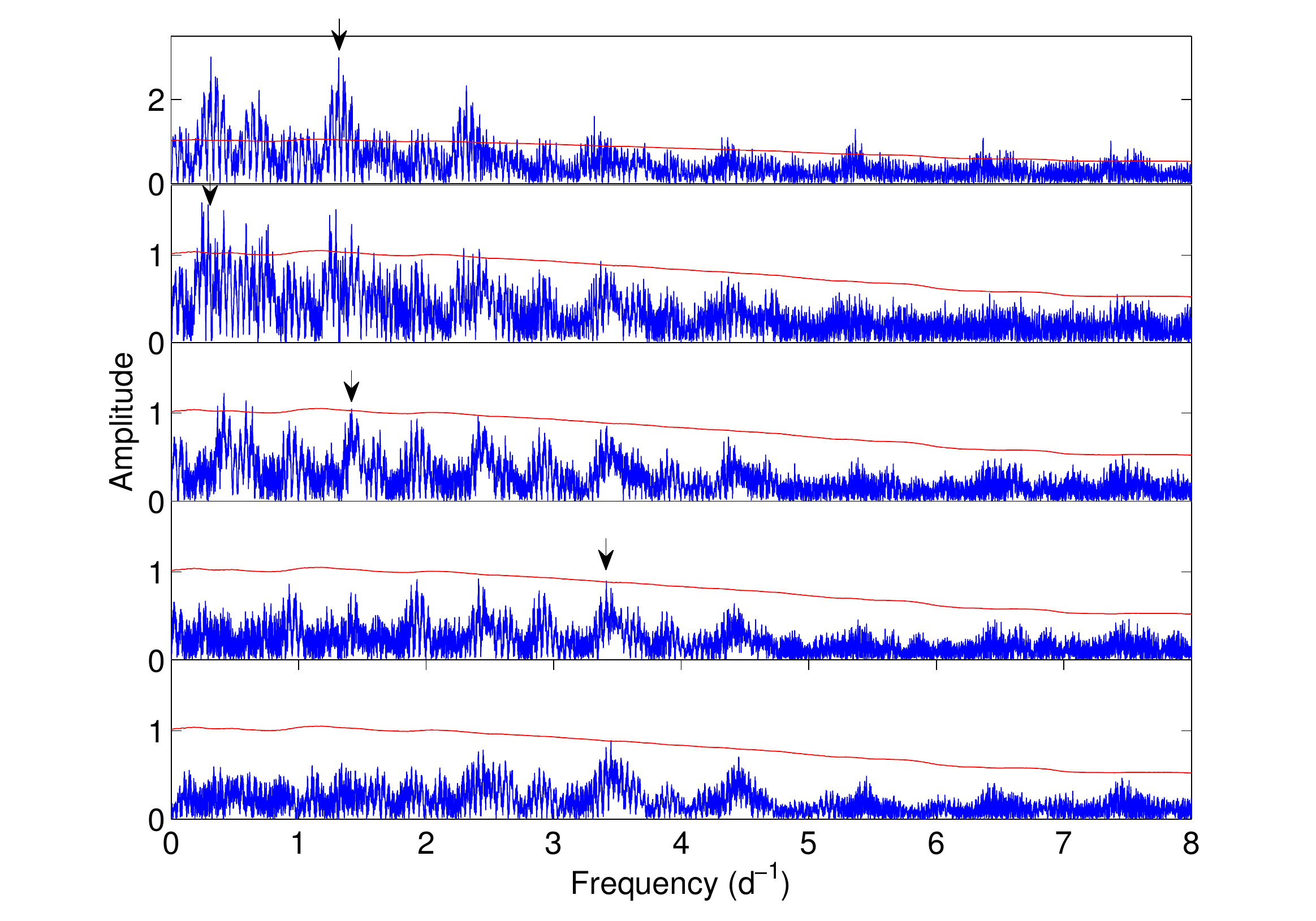}
\label{1momfreqb}
}
\subfigure[]{
\includegraphics[width=0.4\textwidth]{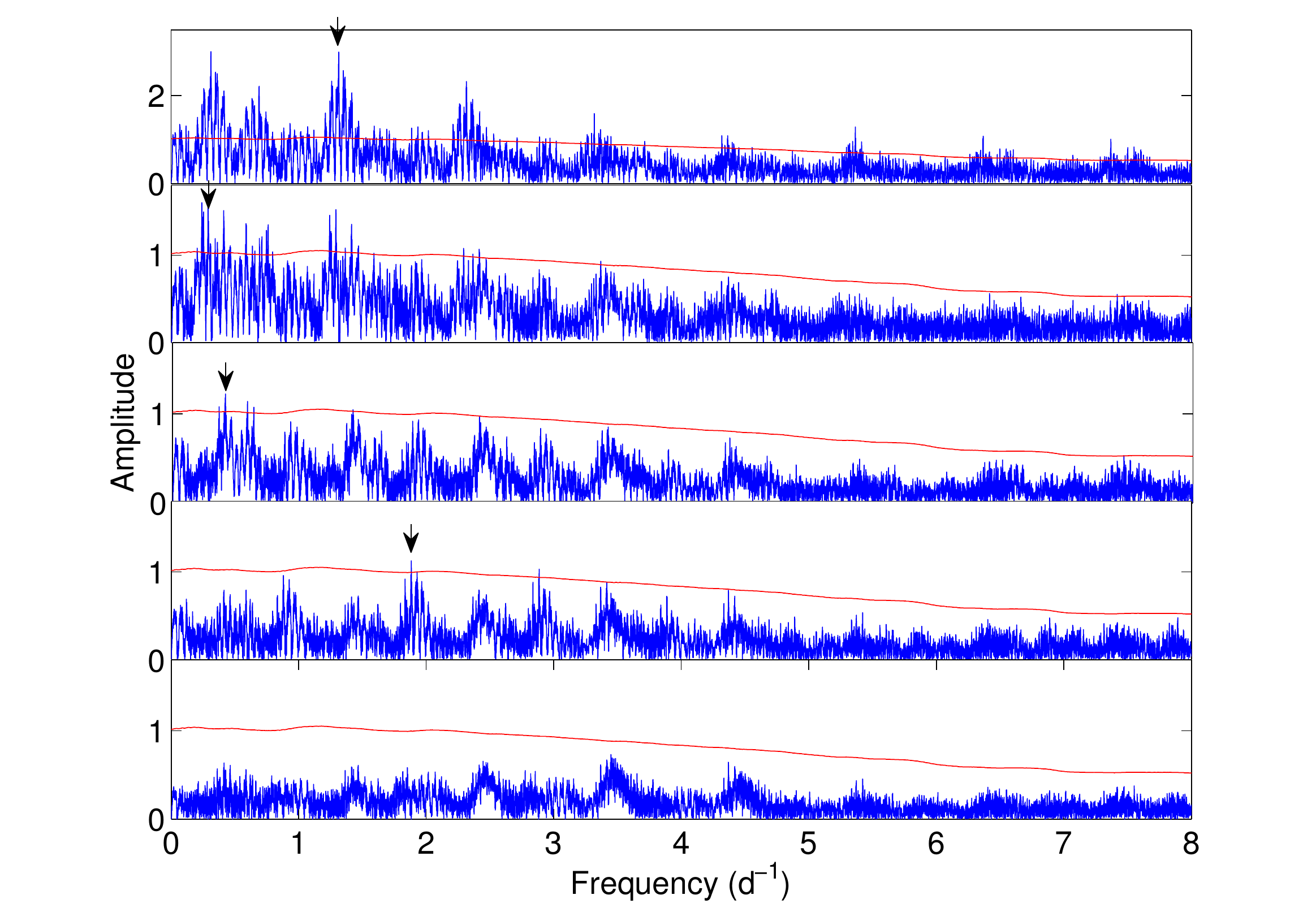}
\label{1momfreqc}
}
\label{1momfreq}
\caption[First Moment Fourier Spectra]{The Fourier spectra showing frequencies
found using the $1^{st}$ moment technique with successive pre-whitening. The red line shows the significance level taken to be the limit of detection.
Path C was taken to be the best selection of frequencies due to the improvement in the percentage variation and overall frequency amplitude.}
\end{figure}

The radial velocity measure is a good frequency analysis tool as it is very sensitive to
any low-degree modes. First moment results are in
good agreement with the results from the pixel-by-pixel search. The Fourier
spectra for the frequencies ($f_a$) are given in Figures \ref{1momfreqa}-\ref{1momfreqc}
and numerical
results are shown in Table \ref{1momres}. Due to an uncertainty in the selection of
the Fourier peaks, three paths of frequency were selected, A, B and C. The first
frequency identified, $0.3122$~d$^{-1}$,  was a 1-day alias of the main
frequency discovered in
all other methods. The second peak, $1.3150$~d$^{-1}$, had an amplitude almost
identical to the first peak and the latter was taken as the true first frequency
for analysis. Increasing amplitudes of Fourier spectra is a known effect of higher noise levels in the $0$ to $1$
d$^{-1}$ region of the Fourier spectrum, which can push an alias peak of a
frequency higher than the real frequency.

There is an ambiguity in the
selection of the  second and third frequencies. There are two possibilities for
the second frequency, either $0.2412$~d$^{-1}$ or $0.2902$~d$^{-1}$.
The $0.2902$~d$^{-1}$ frequency was chosen as the other is not consistent with
the pixel-by-pixel and SigSpec results. The $0.2902$~d$^{-1}$ peak is the
second highest peak in the pixel-by-pixel
method. Pre-whitening using the $0.2902$~d$^{-1}$ frequency removed this
peak entirely from the Fourier spectrum in both methods so it was regarded as
the best choice for $f_{a2}$. The $0.2412$~d$^{-1}$ frequency was still considered
in
the analysis of Path A. The third frequency is more complex as it first appears
as $1.4045$~d$^{-1}$ in the pixel-by-pixel method but then a very close
frequency, $1.4176$~d$^{-1}$, appears using the radial velocity method. It is
worth
noting that in each of the Fourier spectra the other frequency appears strongly
as an additional peak and that choosing the same frequencies as found using the
pixel-by-pixel method gives the same frequency set. When
$1.8831$~d$^{-1}$ was found using this method it had a higher amplitude relative
to $f_{a1}$ than $f_{a4}$ given by choosing just the highest peaks. This confirms our
choice of Path C. 

\begin{figure}
   \includegraphics[width=0.5\textwidth]{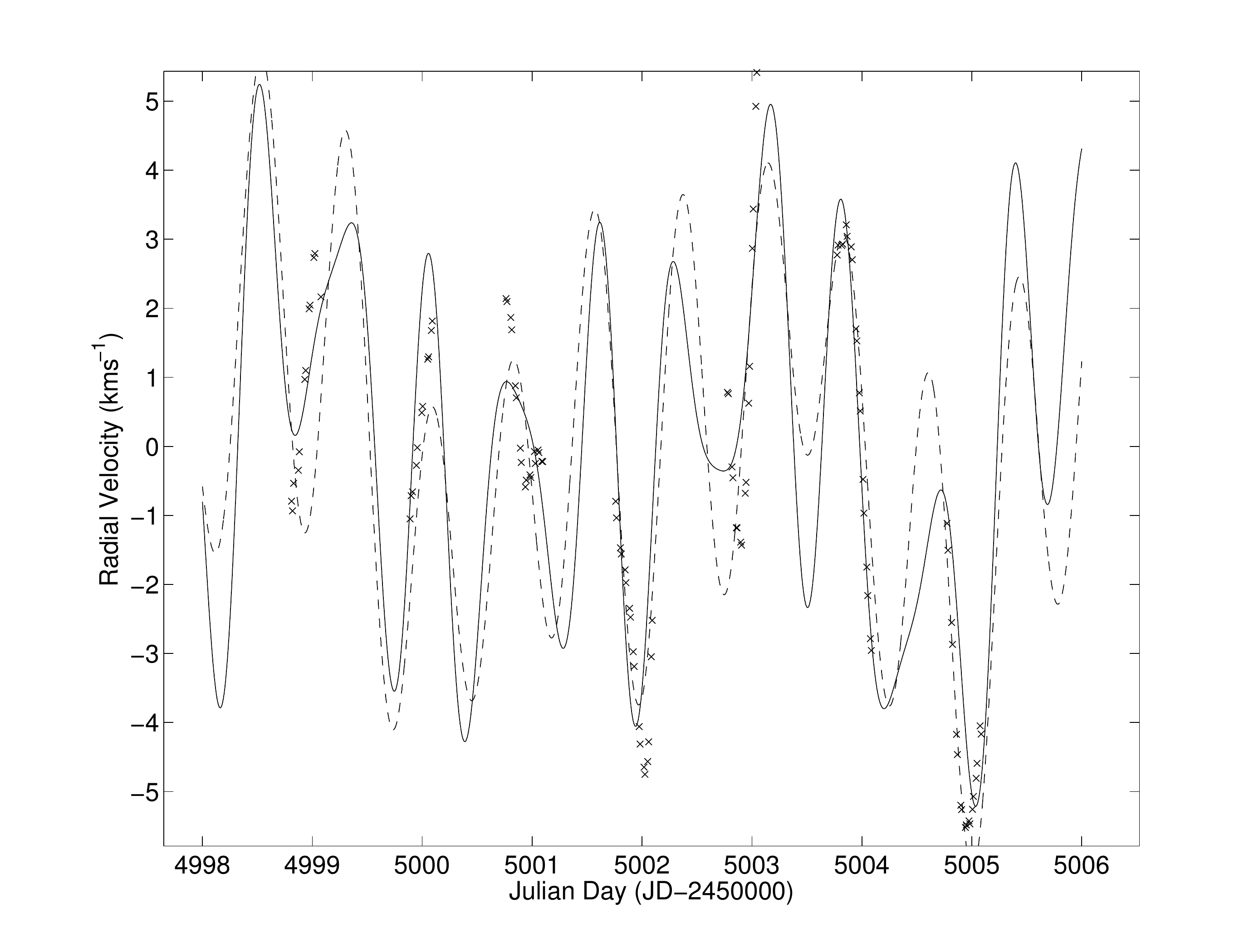}\caption{Radial velocity measurements and fits in a section of the
HD\,135825 data. The crosses show the observed radial velocity and the solid and dashed lines show the fits using Path A and Path
C frequencies respectively.}\label{radvelfit}
\end{figure}

The radial velocity data is plotted over a selected range of Julian date in Figure \ref{radvelfit}. This Figure gives an indication of the difference
in fit
between Path A and Path C.

Even after the removal of four frequencies there still appear to be peaks and
aliases in
the Fourier spectrum that have amplitudes that are too low to significantly distinguish them from the noise. This
could indicate further frequencies are present or that one or more frequencies
have been slightly misidentified.

Table \ref{1momres} shows the amount of variation explained by each frequency
combination.

\begin{center}
\begin{table*}\caption{Frequencies and amplitudes, relative to $f_1a$, $f_1b$, $f_1c$
identified using the variation of the $1^{st}$, $2^{nd}$ and $3^{rd}$ moments respectively. Note (n) identifies the rank of the frequency peak
(high-low) in the
case where the highest peak was not chosen. Paths indicated in bold indicate the best choice of frequencies. See text for further
explanations of the frequency paths.}\label{1momres}
\begin{center}
\begin{tabular}{c | c c c  c c c  c c c}

\multicolumn{10}{c}{First Moment}\\
\hline
ID & \multicolumn{3}{c}{Frequency (d$^{-1}$)} & \multicolumn{3}{c}{Amplitude } & \multicolumn{3}{c}{Variation Explained}\\
\hline
& Path A & Path B & \textbf{Path C} & Path A & Path B & \textbf{Path C} & Path A & Path B &  \textbf{Path C}\\
 \hline
 $f_{a1}$ & \multicolumn{3}{c}{\textbf{1.3150}} & \multicolumn{3}{c}{\textbf{1}} &\multicolumn{3}{c}{\textbf{53\%}}\\
 $f_{a2}$ & 0.2412 & \multicolumn{2}{c}{\textbf{0.2902}(2)} & 0.5351 &\multicolumn{2}{c}{\textbf{0.5278}}& 69\% &\multicolumn{2}{c}{\textbf{67\%}}\\
 $f_{a3}$ & 1.4124 & 1.4149 & \textbf{0.4040}(7) & 0.3931 & 0.4092 & \textbf{0.3388}& 76\% & 77\% & \textbf{75\%}\\
 $f_{a4}$ & 0.9788 & 2.4105 & \textbf{1.8831} & 0.3206 & 0.3085 &\textbf{0.3761}& 83\% & 84\% & \textbf{83\%}\\
\hline 
\multicolumn{10}{c}{  }\\
\multicolumn{10}{c}{Second Moment}\\
 \hline
& Path D & Path E && Path D & Path E && Path D & Path E &\\
 \hline
 $f_{b1}$ & \multicolumn{2}{c}{\textbf{1.3149}} && \multicolumn{2}{c}{\textbf{1}} && \multicolumn{2}{c}{\textbf{63\%}}&\\
 $f_{b2}$ & 0.4398 & \textbf{0.4043}(2) && 0.3675 & \textbf{0.3421} && 70\% & \textbf{70\%}&\\
 $f_{b3}$ & 1.8829 & \textbf{1.8829} && 0.3401 & \textbf{0.3519} && 78\% & \textbf{78\%}&\\
 $f_{b4}$ & 0.2430 & \textbf{0.7125} && 0.2454 & \textbf{0.2666} && 82\% & \textbf{81\%}&\\
\hline
\multicolumn{10}{c}{  }\\
\multicolumn{10}{c}{Third Moment}\\
 \hline
 & Path F & Path G & Path H & Path F & Path G & Path H & Path F & Path G & Path H\\
 \hline
 $f_{c1}$ & \multicolumn{3}{c}{\textbf{1.3150}} & \multicolumn{3}{c}{\textbf{1}} & \multicolumn{3}{c}{\textbf{58\%}}\\
 $f_{c2}$ & 0.4126 & \multicolumn{2}{c}{\textbf{0.2903}(3)} & 0.4330 & \multicolumn{2}{c}{\textbf{0.3948}} & 69\% &\multicolumn{2}{c}{\textbf{67\%}}\\
 $f_{c3}$ & 0.2551 & 0.4150 & \textbf{0.4040}(8) & 0.3417 & 0.3437 & \textbf{0.3142} & 76\% & 77\% & \textbf{75\%}\\
 $f_{c4}$ & 3.4104 & 2.4104 & \textbf{1.8831} & 0.2546 & 0.2807 & \textbf{0.3429} & 83\% & 84\% & \textbf{83\%}\\
\hline
 
\end{tabular}
\end{center}
\end{table*}
\end{center}

\subsection{Second Moment (Variance)}

The second moment reflects variations in the width of
the line profile and is particularly sensitive to even values of $m$. In general we expect the first and second moments to be complementary and
together
build a frequency list similar to the pixel-by-pixel findings. In the analysis the first frequency identified was the usual 1.3149~d$^{-1}$ and the
second frequency that was
recovered was $0.4398$~d$^{-1}$. Another high peak was seen at the previously
identified frequency $0.4043$~d$^{-1}$ so both options were studied as a
possible frequency path. Frequencies ($f_b$) are tabulated for both path D and E in Table \ref{1momres}. Path E
found a higher relative amplitude in $f_{b3}$ than Path D. In the $f_{b4}$ spectrum a 1-day alias
of the previously identified $0.290$~d$^{-1}$ is seen as a high peak in both Fourier spectra. Path E most closely
resembles frequencies identified in the pixel-by-pixel method. Table \ref{1momres} also shows the amount of variation explained by each
frequency
combination.

\subsection{Third Moment (Skewness)}

The third moment frequency analysis is known to produce similar frequency sets
to the first moment method. It describes the variation in the skewness of the
line profile. In this study the third moment also produced the most complicated
frequency spectra with three paths (Path F,G and H) or frequency ($f_c$) sets being identified (see Table \ref{1momres}). This complication
was
due to the ambiguity in peaks in the identification of $f_{c2}$ and $f_{c3}$. In this
study Path H is chosen as that which identifies the same frequencies as the
pixel-by-pixel method and had the highest relative amplitude of $f_{c4}$. The
frequency $0.41$~d$^{-1}$ was found preferentially in $f_{c2}$ and $f_{c3}$ but
vanishes if $0.405$~d$^{-1}$ is chosen. The $0.405$~d$^{-1}$ is still present in
Path F and G as it (or a 1-day alias) occurs as $f_{c4}$ and $f_{c3}$ respectively. This
supports the choice of Path H as the best set of frequencies. Table \ref{1momres} shows the amount of variation explained by each extracted frequency
set.

\subsection{SigSpec Results}

The software package SigSpec \citep{2007AandA...467.1353R} was used to
analyse the two-dimensional moments for frequency identification and significance. SigSpec searches the discrete Fourier transform for
frequencies and assigns spectral significance. Spectral significance is a measure of false-alarm probability for a given
frequency. This probability refers to the likelihood that random noise in the time domain could generate a peak in the Fourier amplitude of similar
size as that of the data itself. For example if the risk a noise peak appears at this amplitude is 1:100\,000 then the false-alarm probability is
0.00001. This value is equivalent to a spectral significance of 5.0 \citep{2007AandA...467.1353R}. The default threshold
value for determining frequencies in SigSpec is set at 5.46 (equivalent signal-to-noise = 4)  but for the purposes of this analysis we took a much
larger threshold of 10. This caution reflects
the requirement of precise frequency
determination for accurate mode identification. For the
three-dimensional pixel-by-pixel dataset, all 48 pixels were individually
analysed for frequency identification using SigSpec. The results for each moment
are given
in Table \ref{sigspecf} for all frequencies ($f_s$) with spectral significance greater than 10. Many of the frequencies found with
significances between
10-15 are not reproduced consistently in other methods of frequency determination and these may not represent true frequencies in the stellar
pulsation.

\begin{center}
\begin{figure}
   \includegraphics[width=0.49\textwidth]{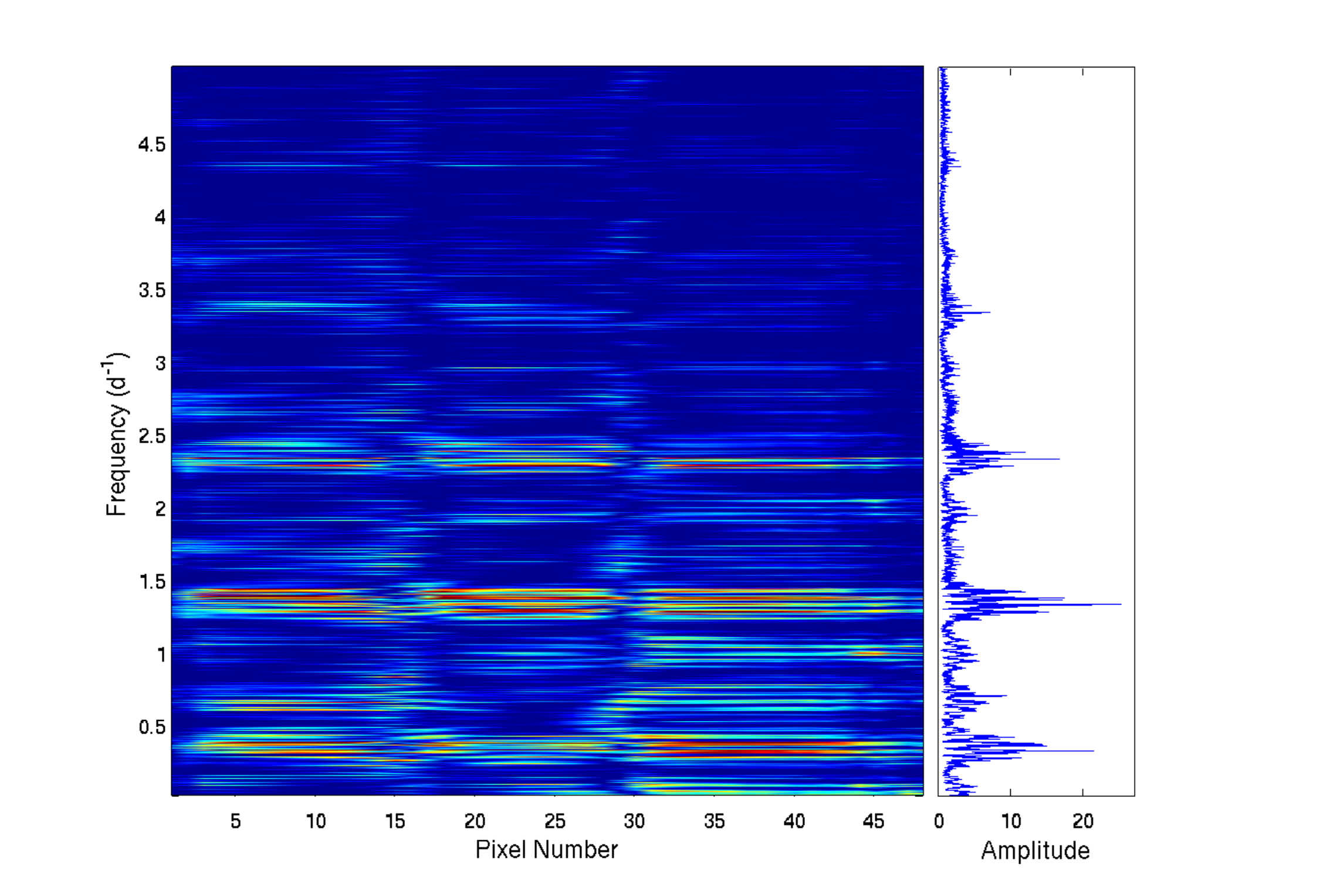}\caption{The
2D peridogram (left) showing frequencies found across each pixel (lighter colours for stronger peaks). The peaks at
1.3~$d^{-1}$and 1-day aliases of this frequency are clearly visible. The mean frequencies of the 48 pixels is plotted on the right.}\label{sigfour}
\end{figure}
\end{center}

Figure \ref{sigfour} shows the Fourier frequency identification across all 48
pixels and shows the three-bump structure of the line-profile variation and the
dominant frequency at $1.315$~d$^{-1}$ and 1-day aliases. 

\begin{center}
\begin{table}\caption[Frequencies and significances found using
SigSpec]{Frequencies found using
SigSpec above the significance threshold of 10. Frequencies that appear in multiple analysis methods are indicated in
bold.}\label{sigspecf}
\begin{center}
\begin{tabular}{ccccccc}
\hline
× & $1^{st}$ mom & Sig & $2^{nd}$ mom & Sig & $3^{rd}$ mom & Sig\\
\hline
$f_{s1}$	&	\textbf{1.3151}	&	33.8	&	\textbf{1.3151}	&	40.4	&\textbf{1.3151}	&	37.9	\\
$f_{s2}$	&	\textbf{0.2906}	&	19.1	&	\textbf{1.4425}	&	13.2	&\textbf{0.4125}	&	16.1	\\
$f_{s3}$	&	\textbf{0.4149}	&	17.0	&	\textbf{1.8834}	&	15.3	&\textbf{0.2541}	&	14.2	\\
$f_{s4}$	&	1.9262	&	15.4	&	0.2435	&	11.8	&\textbf{2.4076}	&	12.5	\\
$f_{s5}$	&	\textbf{1.4457}	&	15.2	&		&		&1.9685	&	12.7	\\
$f_{s6}$	&	2.5257	&	11.9	&		&		&2.4720	&	10.5	\\
$f_{s7}$	&	2.1490	&	11.6	&		&		&2.1855	&	12.3	\\
$f_{s8}$	&		&		&		&		&2.6724	&	10.5	\\
$f_{s9}$	&		&		&		&		&0.0283	&	11.7	\\
$f_{s10}$	&		&		&		&		&0.2013	&	11.3	\\
\hline
\end{tabular}
\end{center}  
\end{table}
\end{center}

\subsection{Frequency Aliases and Combinations}

As this analysis produced a significant set of spectroscopic
frequencies for a $\gamma$ Dor star, a full study of the Fourier spectra was
undertaken. The presence of some combinations for particular frequencies led us
to investigate aliasing patterns and possible combinations of true frequencies
which may show up in our data. This was done using the frequencies found using the pixel-by-pixel method although other independent frequencies found
in the moments were considered. 

The $1$~cycle-per-day alias pattern is obvious in our data (Figure \ref{window}) due to
observations being taken from a single site. Aliasing however does not seem
to add significant uncertainties to our frequency
identification as the aliases have significantly lower power than the identified
frequency. The case where this is most evident is the identification of $f_{p3}$.
Mode identification tests of both frequencies showed $1.405$~d$^{-1}$ was a
better fit to the line profile variation than $0.405$~d$^{-1}$. This was not the case for $f_{p2}$ where the
$0.290$~d$^{-1}$ frequency was found to be a better match to the data, despite
being in the longer range of periods for a $\gamma$ Dor star. 

Combinations of frequencies are known to occur inherently in the star. The
phenomenon has previously been taken into account in the analysis of photometry
of $\delta$ Scuti stars such as in \citet{2005AandA...435..955B}. For our set of
four identified frequencies a grid search of combinations including addition and
subtraction of all frequencies, their multiples and their one-day aliases was undertaken. It
was found that there is possibly a link between $f_{p2}$, $f_{p3}$ and $f_{p4}$ as
$(f_{p2}+1)-f_{p3}\approx f_{p4}-1$. It was also
apparent that
the first identified frequency for $f_3$ in both the first and third moment
is not found indicating it could be an independent frequency. The frequency $f_2$ found first in the second moment is
half that of identified frequency $f_{p4}$ (excluding one-day aliasing).
 
\subsection{Synthetic Frequency Identification}\label{synth}

\begin{center}
\begin{figure}
   \includegraphics[width=0.5\textwidth]{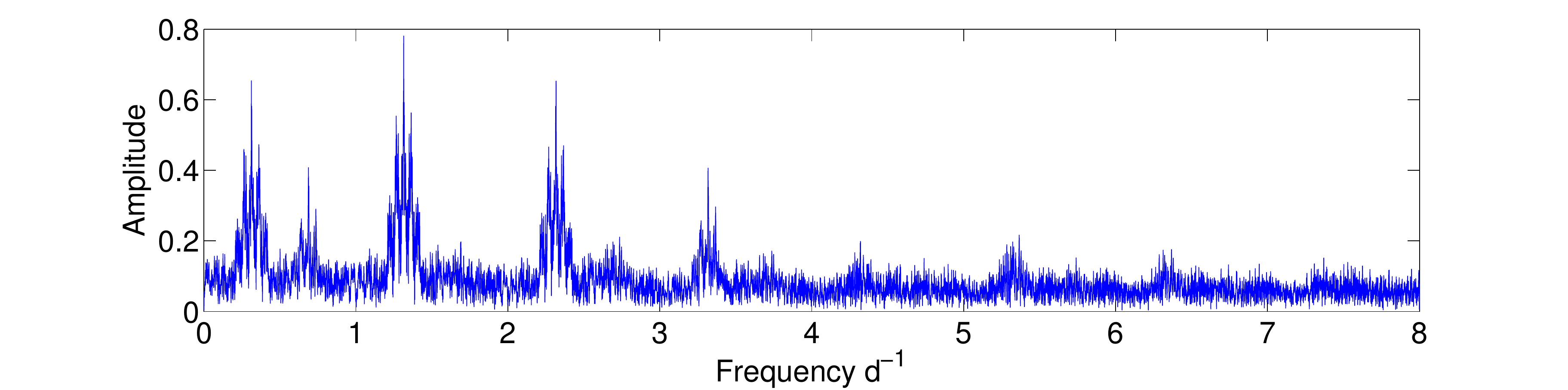}\caption{The Fourier
spectrum of the synthetic data with only one implanted frequency of $f_{p1}=1.3150 $d$^{-1}$ at the observational spacing.}\label{solofreq}
\end{figure}
\end{center}

\begin{center}

\begin{figure}
\includegraphics[width=0.5\textwidth]{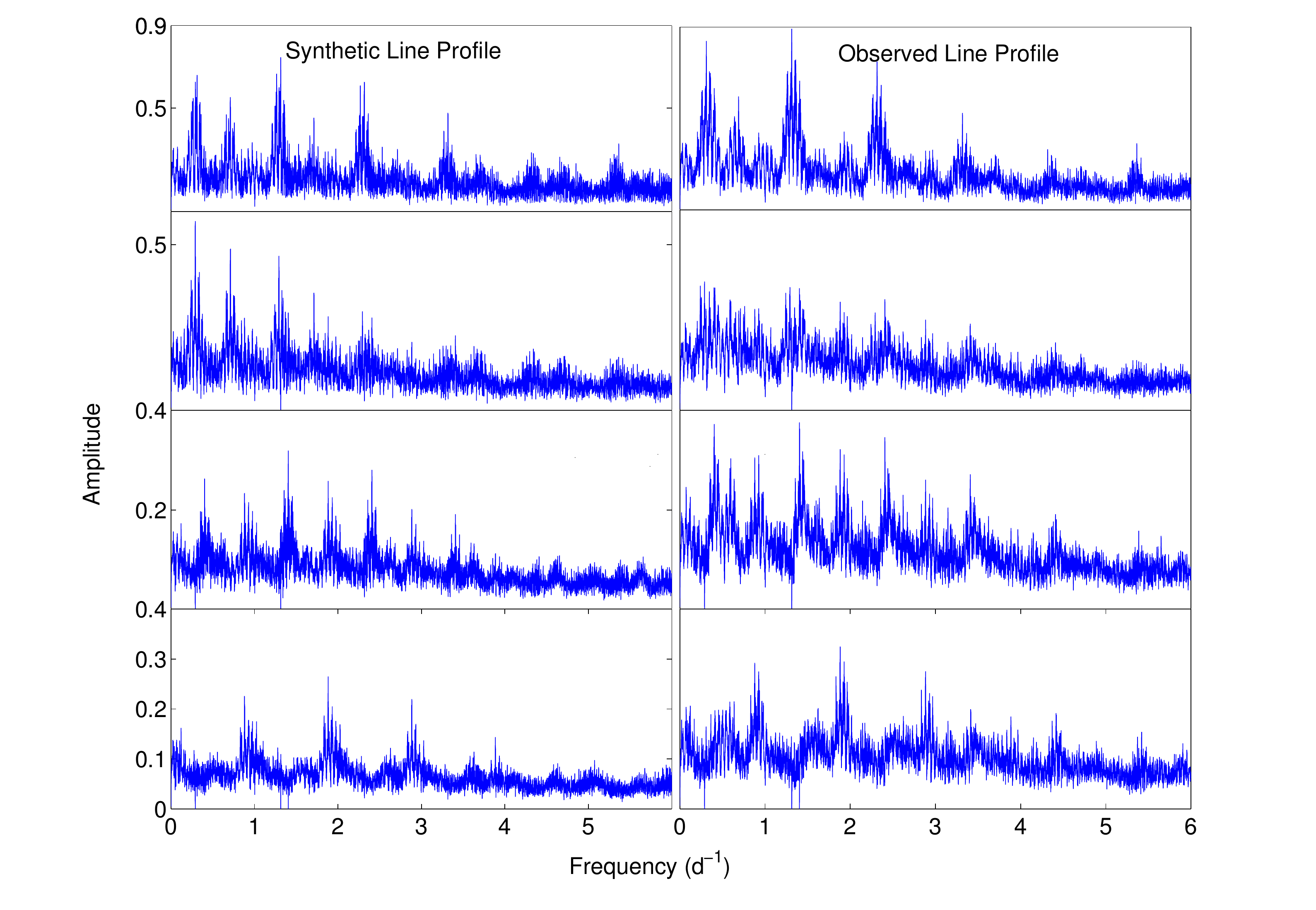}\caption{The
four Fourier spectra for the identified frequencies in the synthetic reproduction (left) and in the real pixel-by-pixel dataset on the
right.}\label{allsynfreq}
\end{figure}
\end{center}

As an additional test of the validity of the derived frequencies a synthetic
data set was constructed using FAMIAS \citep{2008CoAst.157..387Z}. The data
used the same time spacings as
the real observations and then the four frequencies, their amplitudes, phases
and modes (as determined in Section \ref{modeid}) were entered along
with the known parameters of the star and representative spectral line. The
results for the derivation of the first frequency alone in the data and the
first four frequencies together are given in Figures \ref{solofreq} and
\ref{allsynfreq} respectively. All frequencies implanted in the line profiles
were able to be extracted, although $f_{p4}$ appeared in the Fourier spectra as a
higher peak than $f_{p3}$. The frequencies $2.80$~d$^{-1}$ and $1.11$~d$^{-1}$ are
found as the fifth and sixth frequencies which indicates they are artefact
frequencies as only the above four frequencies inserted should be recovered in the Fourier spectrum.


\subsection{Frequency Results}\label{freqres}

A summary of the frequencies found by all methods is presented in Table
\ref{allfreq}. An error estimate for each frequency ($\sigma(f)$) is given using the definition of \cite{2008AandA...481..571K} who propose:

\begin{equation}
\sigma(f) = \frac{1}{T*\sqrt{sig(a)}},
\end{equation}where T is the time base of observations in days and $sig(a)$ is the spectral significance of the frequency from SigSpec. This equation
provides a low
uncertainty on our frequencies due to the long time base of observations, but should be regarded as a guide only as the authors
feel this uncertainty under-represents the errors present in the frequency identification. 

A further least-squares fit of the frequencies to the one-dimensional data was performed to get a second estimate on frequency errors by
examination
of the 95\% confidence bounds. The bounds gave errors on the order of $\pm0.00002$ d$^{-1}$. These errors were disregarded and the previously found
SigSpec uncertainties adopted from this point as they are an order of magnitude larger than the confidence fits and it is prudent to remain cautious
about the precision of these results.

Four frequencies are confirmed in this study: $1.3150\pm 0.0003$~d$^{-1}$,
$0.2902\pm 0.0004$~d$^{-1}$, $1.4045\pm 0.0005$~d$^{-1}$ and $1.8829\pm 0.0005$~d$^{-1}$ (Table
\ref{allfreq}) These are the frequencies found originally in the pixel-by-pixel measurements and confirmed in the $1^{st}$ -- $3^{rd}$ moment
analysis with uncertainties calculated from significances in the $1^{st}$ moment. Additional frequencies that may be present in the data are
$1.44$~d$^{-1}$, $0.25$~d$^{-1}$ and $1.415$~d$^{-1}$. These frequencies had either poor signal-to-noise or misshapen amplitude variation profiles
which limited any further mode-identification.

From the identification of multiple frequencies in the 0.3-3 d$^{-1}$ range we can confirm HD\,135825 is a true $\gamma$ Dor star. After the
removal of four frequencies in all the above methods there remained some indication of a periodic signal. It is likely there are further $\gamma$
Dor frequencies present in the data below our detection limit with this dataset. There do not
appear to be any higher frequencies suggestive of a $\delta$ Scuti hybrid. 

\begin{center}
\begin{table*}\caption[Frequency Results for all Methods]{Frequencies (in d$^{-1}$) identified
using pixel-by-pixel (pbp) and $0^{th}$ to $3^{rd}$ moment methods. Note (a) denotes where a one-cycle-per-day alias was
found as the highest peak and (\#) identifies frequencies which did not have the
highest peak but were sequenced according to the number in brackets and chosen
as a better result.}\label{allfreq}
\begin{center}
\begin{tabular}{lllllll}
\hline
 & pbp & 0th  & 1st  & 2nd  & 3rd  & final (d$^{-1}$) \\
 \hline
$f_1$ & 1.3150 & 1.3147 & 1.3122 (a) & 1.3149 & 1.3150  & 1.3150 $\pm$ 0.0003\\
$f_2$ & 0.2902 & × & 0.2902 (2) & 0.4043 (2) & 0.2903 (3) & 0.2902 $\pm$ 0.0004\\
$f_3$ & 1.4045 & × & 0.4039 & 1.8829 & 0.4040 (8) & 1.4045 $\pm$ 0.0005\\
$f_4$ & 1.8829 & × & 1.8831 & × & 1.8831 & 1.8829 $\pm$ 0.0005\\
\hline
\end{tabular}
\end{center}
\end{table*}
\end{center}

\section{Mode Identification}\label{modeid}

The first calculation performed was a fit to the zero-point profile using the parameters $v$sin$i$, equivalent width and velocity offset.
The latter is used to bring the observational data onto a consistent scale. These
values were then used to refine future searches in parameter space. To continue with mode identification of the pulsation frequencies, the
three Fourier parameters (zero-point, amplitude and phase variations across the
cross-correlated line profiles) were calculated. Using the Fourier Parameter Fit
Method \citep{2009AandA...497..827Z}, the modes of each of the final frequencies in
Table \ref{allfreq} were found independently, as well as the combination of the
four frequencies. This was done using the frequency analysis package in FAMIAS
\citep{2008CoAst.157..387Z}. The fits for each mode were performed on the amplitude and phase distributions and a $\chi^{2}$ calculated.  The
pulsation parameter space that was searched (using a genetic algorithm optimisation to search parameter space) is outlined in Table
\ref{param}. 

\begin{figure*}
\centering
\subfigure[$f_1 = 1.3150$~d$^{-1}$ fit with (1,1) mode, $\chi^{2} = 12.9$.]{  
\includegraphics[width=0.45\textwidth]{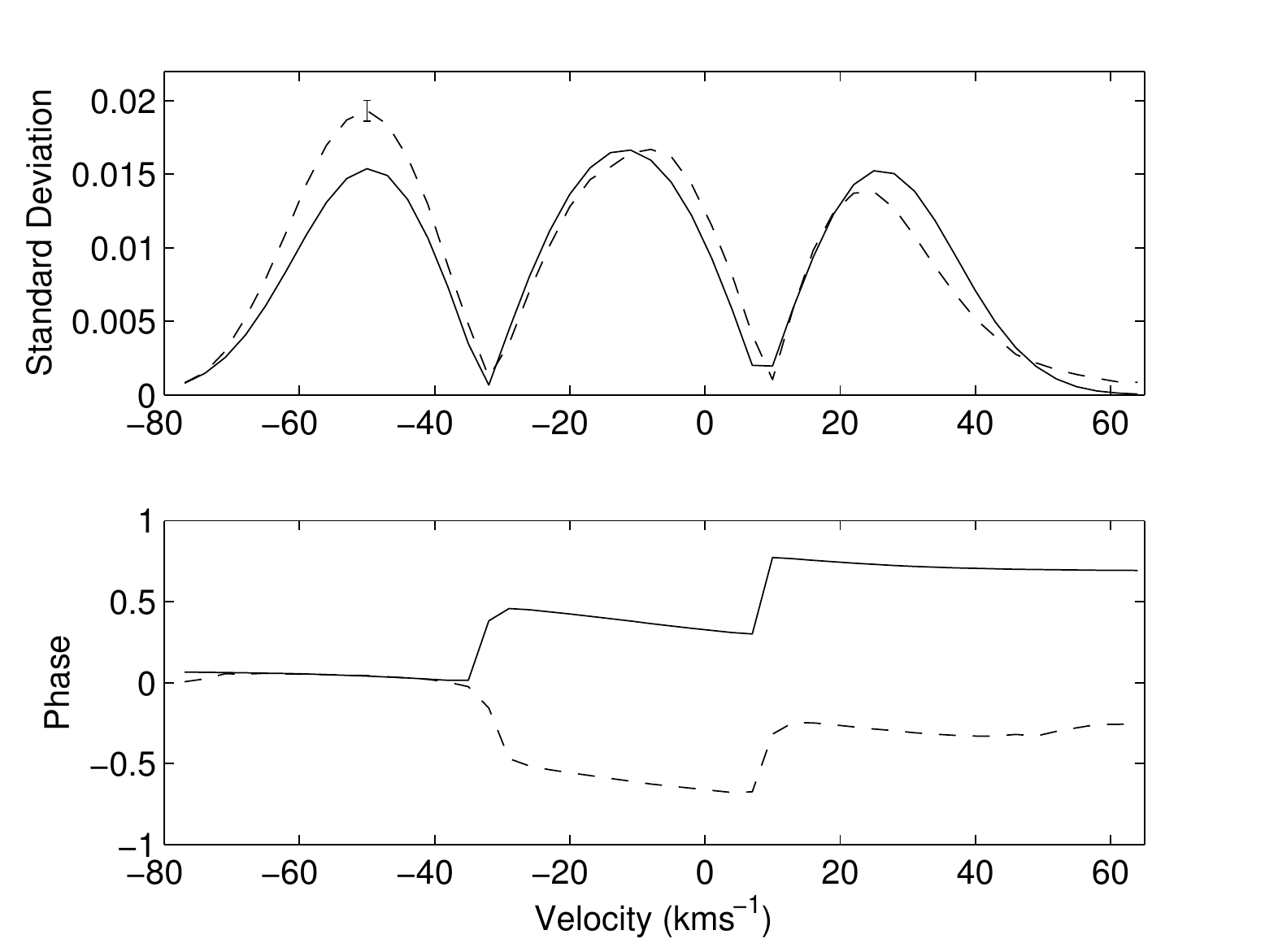}\label{modef1}
}
\subfigure[$f_2 = 0.2902$~d$^{-1}$ fit with (2,-2) mode, $\chi^{2} = 3.1$.]{  
\includegraphics[width=0.45\textwidth]{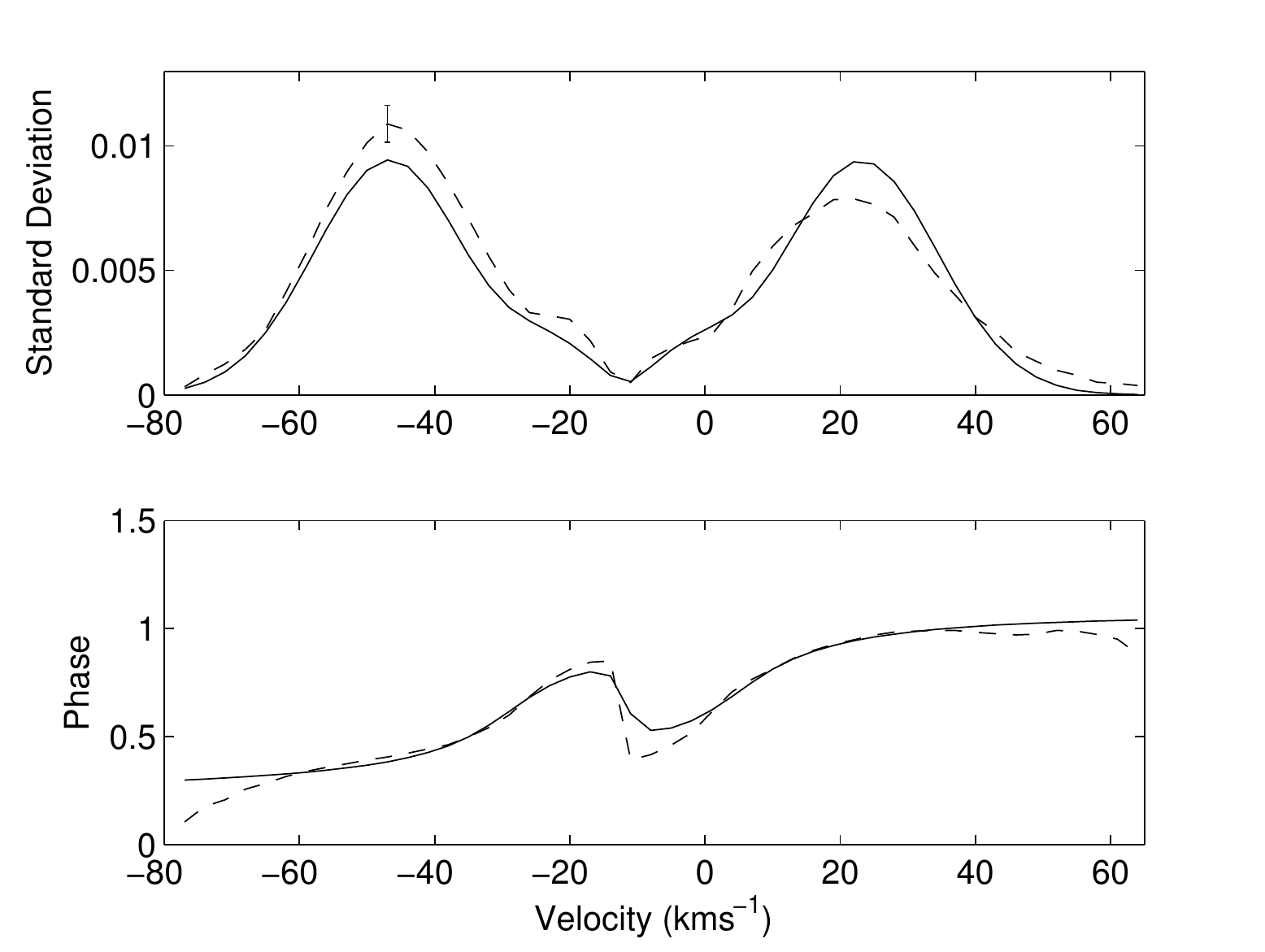}\label{modef2}
}
\subfigure[$f_3 = 1.4045$~d$^{-1}$ fit with (4,0) mode, $\chi^{2} = 10.6$.]{  
\includegraphics[width=0.45\textwidth]{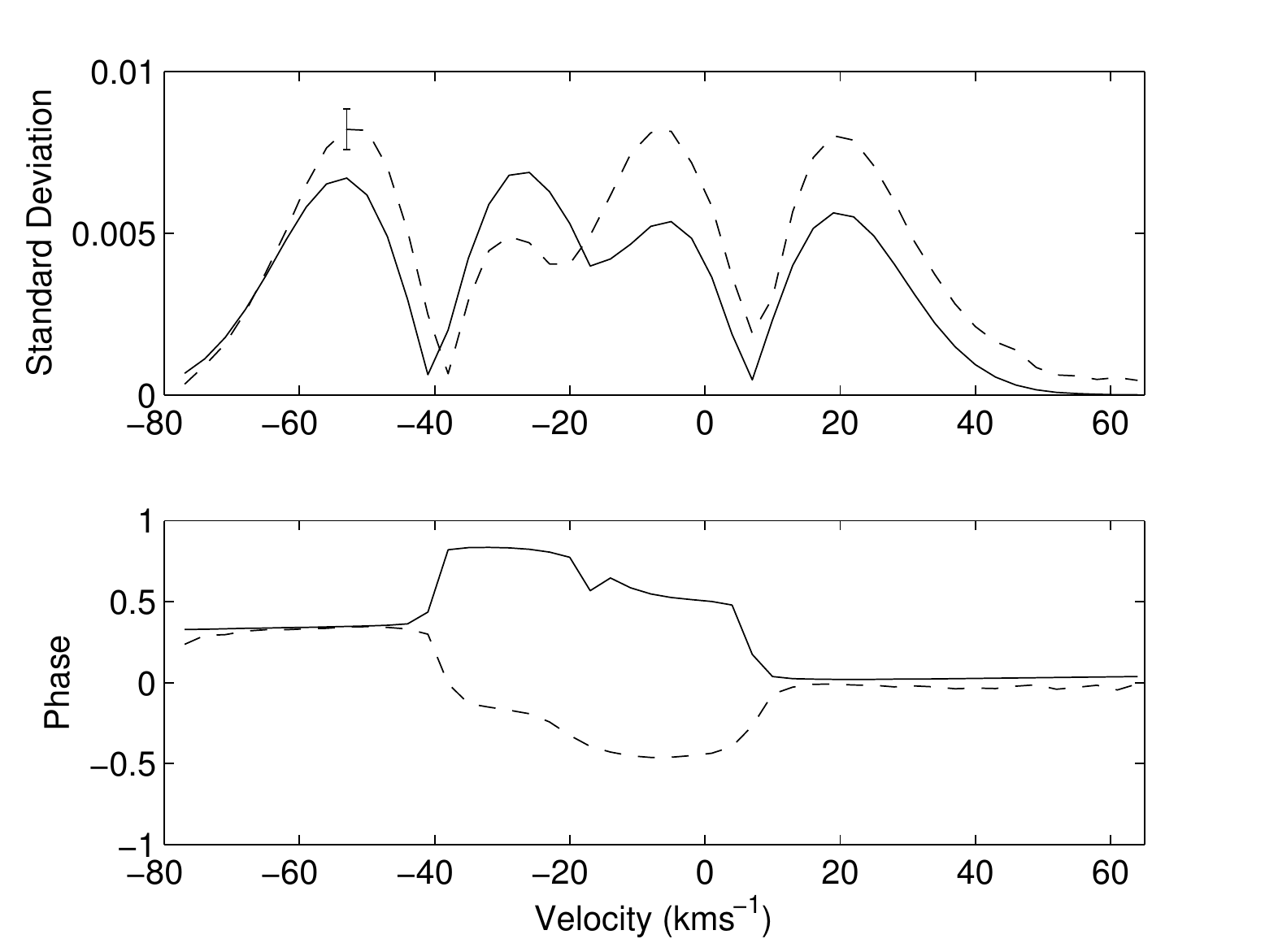}\label{modef3}
}
\subfigure[$f_4 = 1.8829$~d$^{-1}$ fit with (1,1) mode, $\chi^{2} = 2.5$.]{  
\includegraphics[width=0.45\textwidth]{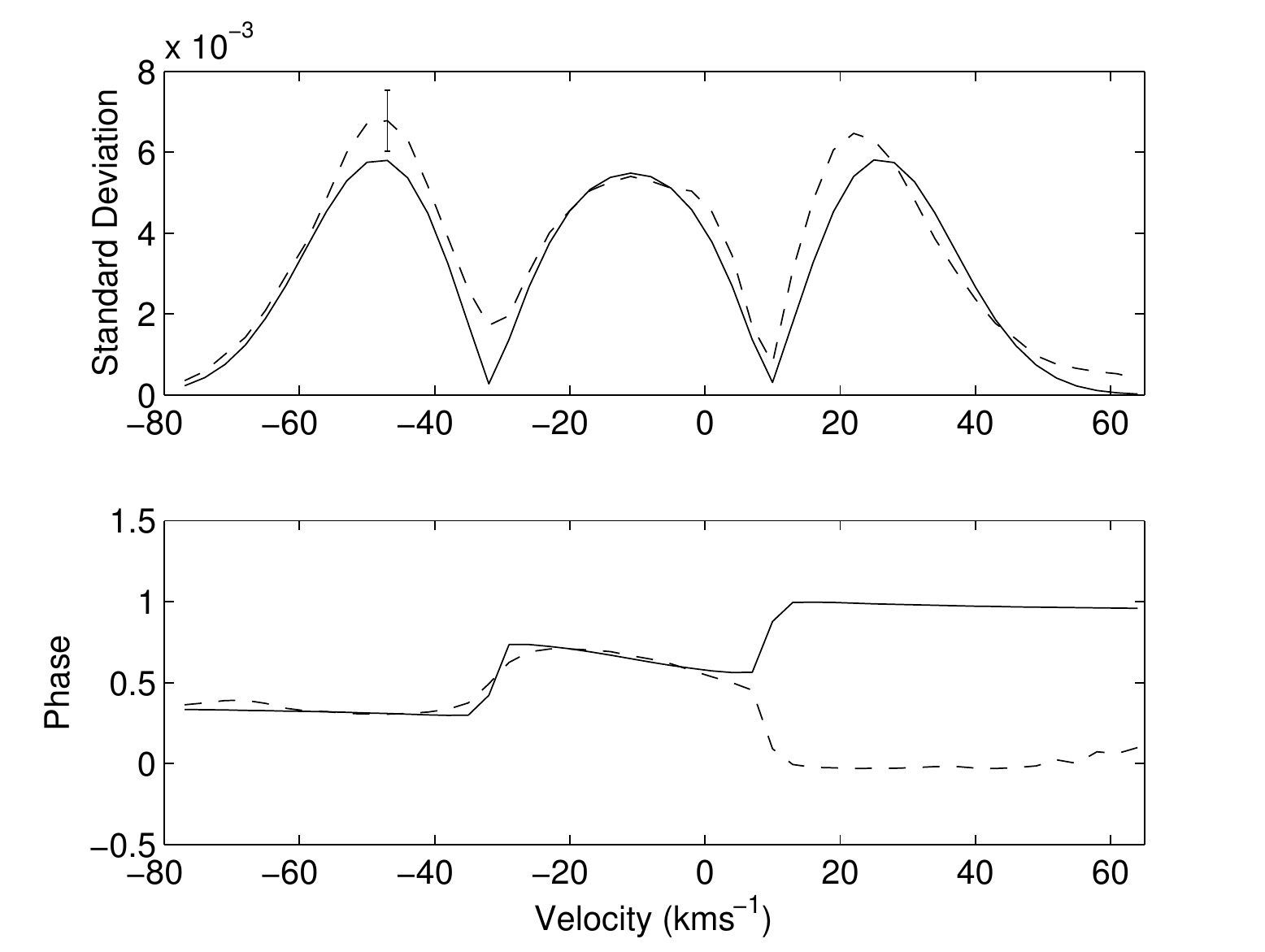}\label{modef4}
}
\caption{Mode identification results. Amplitude variation and
phase across the line profile (solid) of the four identified frequencies with the lowest $\chi^{2}$ fits (dashed). A maximum error bar for the fits
is given in each amplitude variation plot. Errors on the data and phases are too small to display.}
\end{figure*}

Figures \ref{modef1}-\ref{modef4} show the Fourier parameters and their fits to
the amplitude and phase across the line profile for the
four frequencies. Note that FAMIAS sometimes generates phases (see Figures \ref{modef1},\ref{modef2},\ref{modef4}) that are $\pm$1 cycle ($\pm$2$\pi$)
displaced with respect to the observations. The mode fits are best for the even frequencies, $f_2$ ($=$0.2902 d$^{-1}$; (l,m)$=$(2,-2)) and $f_4$
($=$1.8829 d$^{-1}$; (1,1)) that have $\chi^{2}$
$=$ 3.1 and 2.5 respectively, whereas $f_1$ ($=$1.3150 d$^{-1}$; (1,1)) and $f_3$ ($=$1.4045 d$^{-1}$; (4,0)) have $\chi^{2}$ of 12.9 and 10.6
respectively. The most significant
deviations are in the $f_3$ fits where the amplitude and phase are not fitted well, albeit with lower pulsational amplitudes than $f_1$ and $f_2$.

\begin{figure*}
\centering
\subfigure[\protect$\chi^\protect{2\protect}\protect$ 
values for \protect$l\protect=0-3$ for
$f\protect_1\protect=1.3150$d$\protect^\protect{-1\protect}$.\newline \hspace*{1.5em} Best fit value is for ($l,m$) = (1,1).]{  
\includegraphics[width=0.45\textwidth]{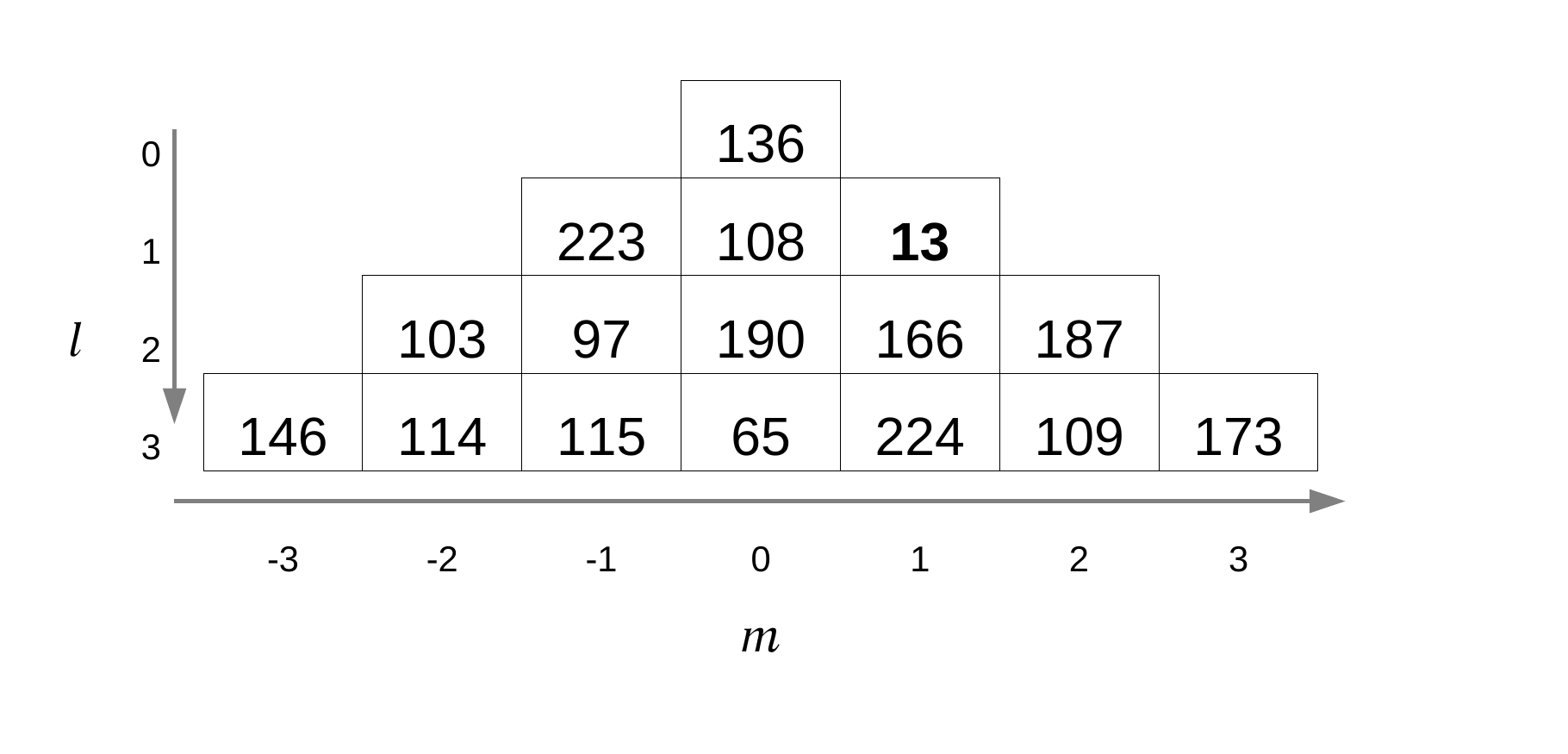}\label{bestfit1}
}
\subfigure[\protect$\chi^\protect{2\protect}\protect$ 
values for \protect$l\protect=0-3$ for
$f\protect_2\protect=0.2902$d$\protect^\protect{-1\protect}$. \newline \hspace*{1.5em}
Best fit value is for ($l,m$) = (2,-2).]{  
\includegraphics[width=0.45\textwidth]{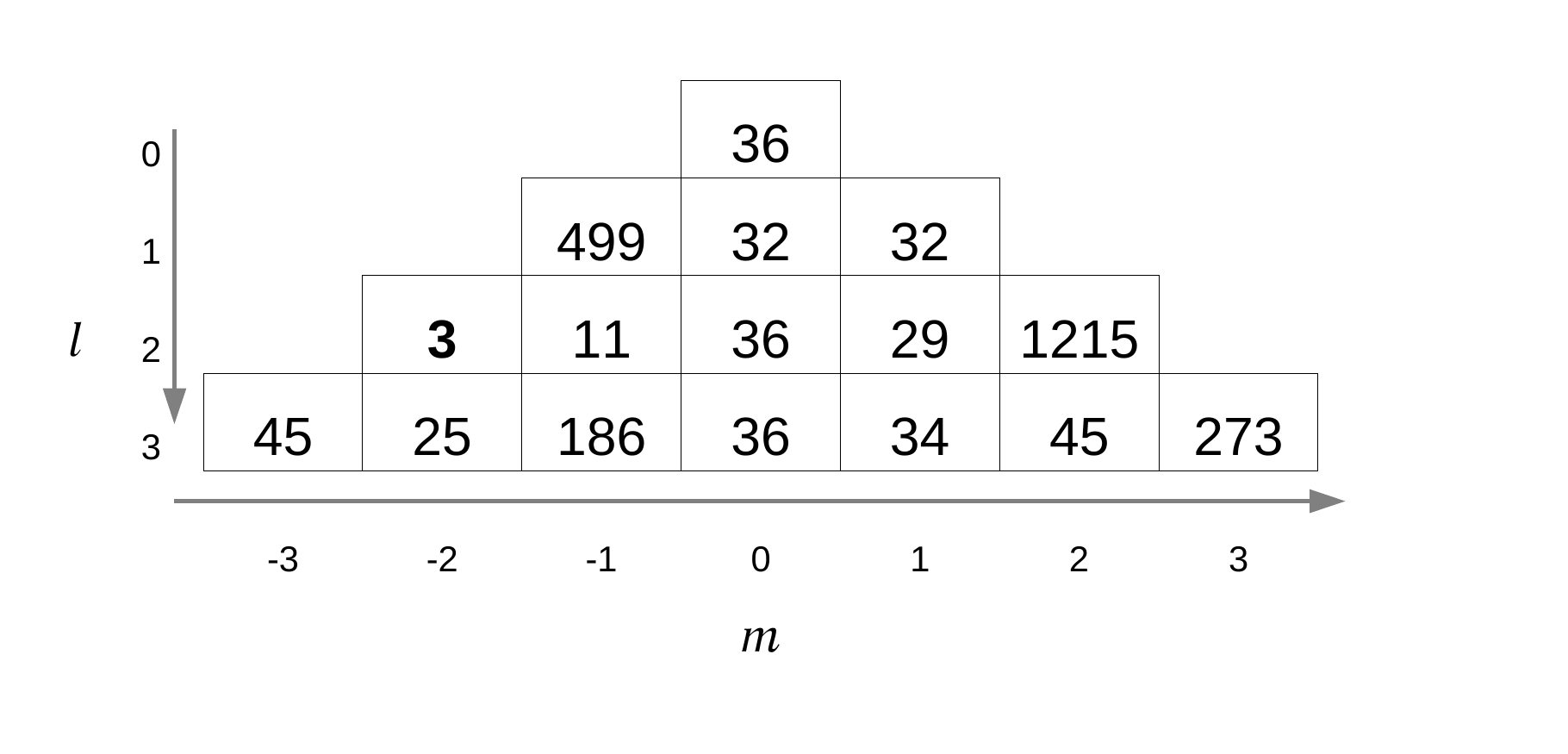}\label{bestfit2}
}
\subfigure[\protect$\chi^\protect{2\protect}\protect$ 
values for \protect$l\protect=0-4$ for
$f\protect_1\protect=1.4045$d$\protect^\protect{-1\protect}$. \newline \hspace*{1.5em}
 Best fit value is for ($l,m$) = (4,0).]{  
\includegraphics[width=0.45\textwidth]{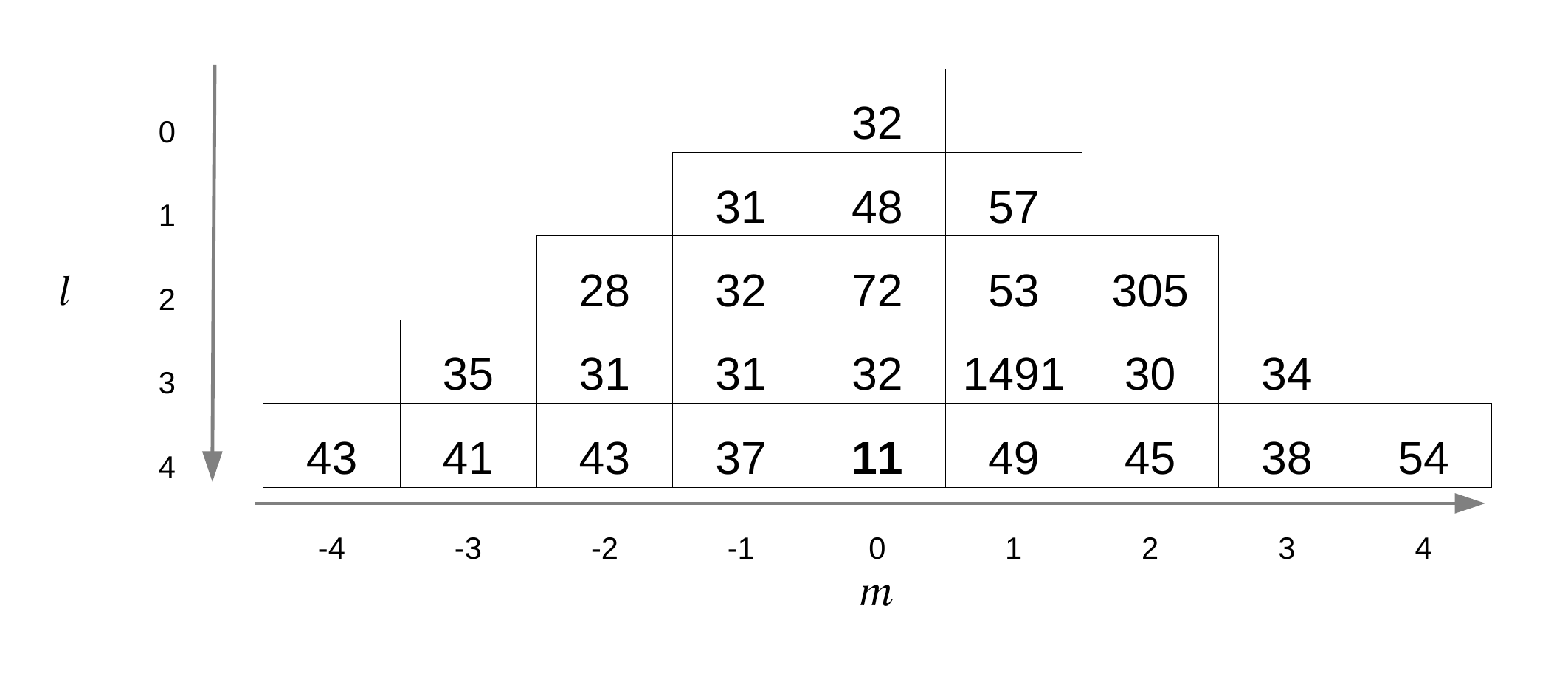}\label{bestfit3}
}
\subfigure[\protect$\chi^\protect{2\protect}\protect$ 
values for \protect$l\protect=0-3$ for $f\protect_1\protect=1.8829$d$\protect^\protect{-1\protect}$. \newline \hspace*{1.5em}
Best fit value is for ($l,m$) = (1,1).]{  
\includegraphics[width=0.45\textwidth]{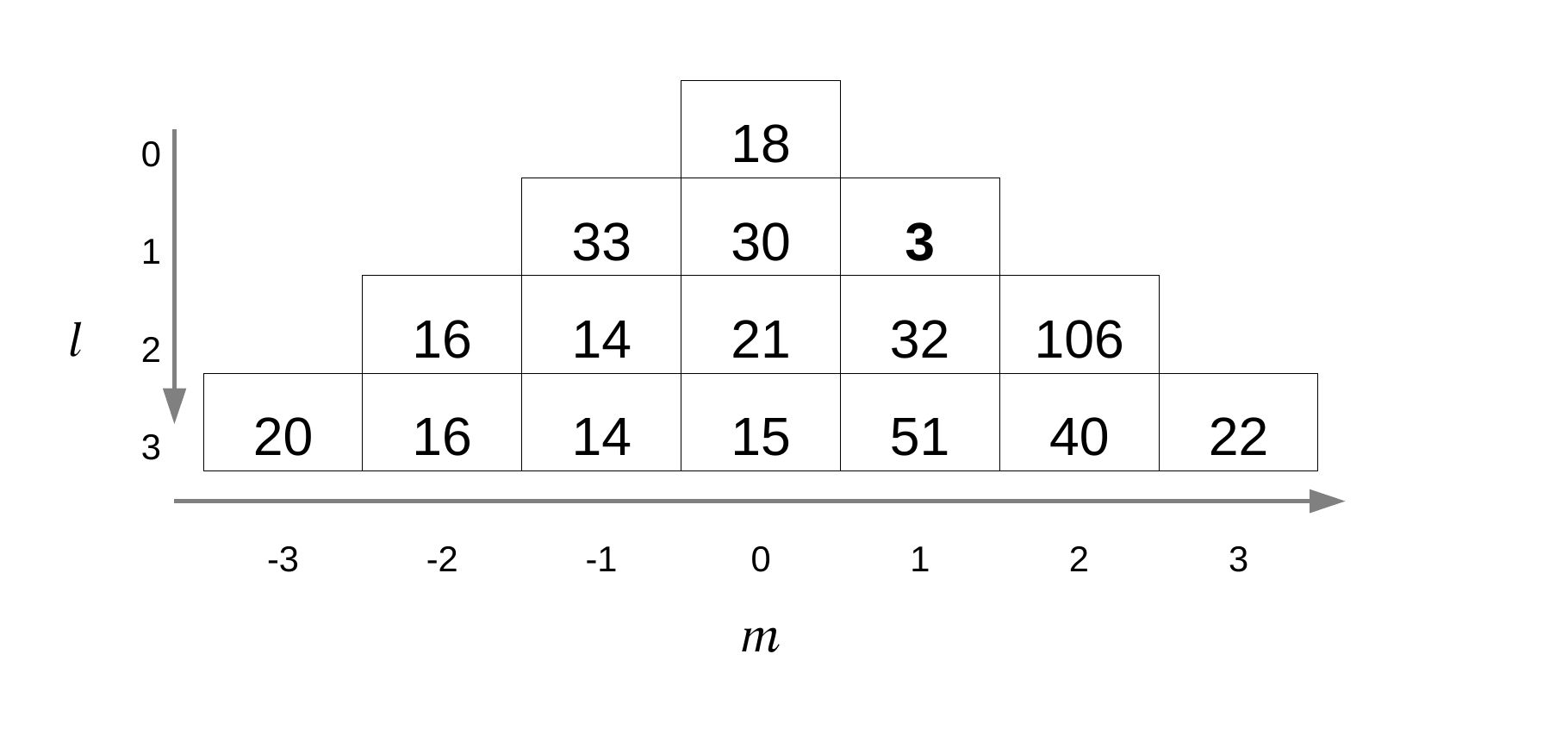}\label{bestfit4}
}
\caption{Lowest \protect$\chi^\protect{2\protect}\protect$ 
values for each possible (\protect$l,m\protect$) combination for the final identified frequencies. The best fit
\protect$\chi^\protect{2\protect}\protect$ is identified in bold.}
\end{figure*}

The final parameters of the fit to each of the modes individually, the
$\chi^{2}$ value and stellar parameters are given in
Table \ref{allmode}. The mode with the lowest $\chi^{2}$ fit was selected as the best mode. This was an unambiguous selection as there was only one
mode for each frequency with a significantly lower $\chi^{2}$. The $\chi^{2}$ of each mode fit to the final frequencies is
given in Figures \ref{bestfit1}-\ref{bestfit4}. The individual $\chi^{2}$ values can be used as a guide to the goodness of fit but frequencies with
higher amplitudes
and small errors have tighter fit constraints. Couple this with the poorly modelled asymmetry in the line profile and the $\chi^{2}$ can be
higher than is expected with a good fit such as for $f_1$. When all modes were combined for a simultaneous fit a reduced
$\chi^{2}$ of $12.7$ was achieved with the stellar variables and line parameters in Table \ref{allmode}. The other best possible fits to the data
are shown in Table
\ref{altmode}. Although some alternate fits to $f_3$ have a similar $\chi^{2}$, the shape of these other fits do not show the four bumps as in the
line profile variation. The second and third best fit modes have three and two bumps respectively in the amplitude variation profiles and do not
display four distinct changes in the phase and thus we are confident the (4,0) identification is the best
match.

The best individual values for the parameters are given by the best fits which
are to the second and fourth frequencies. The values for the first frequency
also merit consideration given its dominant contribution to the
line-profile variation. The range of values found for the inclination of the
star show it is an uncertain result of the search, but values between
$50\degree-90\degree$ are reasonable and an inclination close to $90\degree$ may
explain the number of detected frequencies, as we observe the star edge-on. This means we see the full surface of the star as it rotates. The value
for $v$sin$i$ was measured from the line profile to
be 38 $\pm$ 5 kms$^{-1}$
\citep{2006AandA...449..281D}. 

\begin{center}
\begin{table}\caption[Stellar parameters used in the mode
identification]{Stellar parameters used in the mode
identification. Fixed values for Teff, [M/H] and log $g$ were taken from
\cite{2008AandA...478..487B} and $v$sin$i$ values from \cite{2006AandA...449..281D}. Values for the other parameters were taken
as reasonable limits for $\gamma$ Dor stars. }\label{param}
\begin{center}
\begin{tabular}{lllll}
\hline
Property & Fixed Value & Min & Max & Step\\
\hline
Radius(solar units)  & × & 1 & 5 & 0.1\\
Mass (solar units) & × & 0.5 & 5 & 0.01\\
Temperature (K) & 7050 & × & × & ×\\
Metallicity [M/H] & 0.13 & × & × & ×\\
log $g$ & 4.39 & × & × & ×\\
Inclination ($\degree$) & × & 0 & 90 & 1\\
$v$sin$i$ (kms$^{-1}$) & × & 30 & 50 & 1\\
\hline
\end{tabular}
\end{center}
\end{table}
\end{center}

\begin{center}
\begin{table}\caption[Results of mode identification for four frequencies
individually]{Results of mode identification for $f_1$ alone, all four
frequencies
individually after a least-squares fit is applied (lsf), and all four
frequencies simultaneously (sim).}\label{allmode}
\begin{center}
\begin{tabular}{ccccccc}
\hline
freq & Mode ID & $\chi^{2}$ & Inclination  & Vel. amp & Phase\\
 & &  & ($\degree$) & (kms$^{-1}$)  & \\
\hline
$f_1$ & (1,1) & 7.6 & 87 & 0.78 & 0.64\\
\hline
$f_1$ lsf & (1,1) & 12.9 & 64 & 0.78 & 0.64\\
$f_2$ lsf & (2,-2) & 3.1 & 55 & 0.65 & 0.43\\
$f_3$ lsf & (4,0) & 10.6 & 88 & 0.65 & 0.40\\
$f_4$ lsf & (1,1) & 2.5 & 51 & 0.65 & 0.88\\
\hline
$f_1$ sim & (1,1) & 12.7 & 87 & 0.80 & 0.63\\
$f_2$ sim & (2,-2)& & & 0.50 & 0.41\\
$f_3$ sim & (4,0) & & & 0.67 & 0.40\\
$f_4$ sim & (1,1) & & & 0.69 & 0.91\\
\hline
\end{tabular}
\end{center}
\end{table}
\end{center}

\begin{center}
\begin{table}\caption{Best fits for modes identified for HD\,135825}\label{altmode}
\begin{center}
\begin{tabular}{ccccc}
\hline
$f_{1}$ & $f_{2}$ & $f_{3}$ & $f_{4}$ & $\chi^{2}$\\
\hline
(1,1) & (2,-2) & (4,0) & (1,1) & 10.6\\
 &  & (1,1) &  & 12.7\\
 &  & (2,-2) &  & 13.9\\
 &  & (5,4) &  & 14.9\\
 &  & (3,-1) &  & 15.5\\
 &  & (3,-3) &  & 15.6\\
 &  & (4,3) &  & 15.8\\
\hline
\end{tabular}
\end{center}
\end{table}
\end{center}

\subsection{Rotation and Pulsation Parameters}

FAMIAS has a tool to provide a check for the validity of the mode identification method based on the rotational parameters of the star. The 
stellar rotational parameters of equatorial rotational velocity (v$_{rot}$; v$_{rot}$ = 2$\pi$R$f_{rot}$), rotational period
(T$_{rot}$), rotational frequency
($f_{rot}$),
critical velocity (v$_{crit}$) , critical vsini ($v$sin$i_{crit}$) and critical minimum inclination ($i$) are calculated from inputs of mass, radius,
$v$sin$i$ and inclination using estimates for mass and radius typical of $\gamma$ Dor stars (1.5 M$_\odot$ and 1.6 R$_\odot$) and $v$sin$i$ and
inclination
from the mode identification (39.7 kms$^{-1}$ and 87$\degree$). The results are tabulated below:
\begin{center}
\begin{tabular}{ll}
\\
v$_{rot}$ & 39.75 kms$^{-1}$ \\
T$_{rot}$ & 2.03 d\\
$f_{rot}$ & 0.49 d$^{-1}$\\
v$_{crit}$ & 423 kms$^{-1}$\\
$v$sin$i_{crit}$ & 423 kms$^{-1}$\\
$i_{crit}$ & 5.4$\degree$\\
\end{tabular}
\end{center}
From the above table we can see that HD\,135825
is
not near the break-up limit of the star and is naturally rotating with a period of around two days. The ratio of horizontal to
vertical amplitude of the pulsation ($\kappa$), is calculated using
the equation

\begin{equation}\label{kvalue}
\kappa = \frac{GM}{\omega^{2}R^{3}},
\end{equation}where $G$ is the gravitational constant, $R$ and $M$ are the stellar mass and radius and $\omega$ is the angular frequency. For the
various
frequencies $\kappa$ was much greater than $1.0$ which indicates low-frequency g-modes as we expect for $\gamma$ Dor stars.
We can also use the determined inclination to transform our observed frequencies into the co-rotating frequencies ($f_{co-rot}$) required to model the
star ($f_{co-rot} = f_{obs} - mf_{rot}$). These
frequencies, along with the horizontal-to-vertical amplitude ratios calculated using Equation \ref{kvalue}, are given in Table \ref{stepul}. All of
the frequencies lie now in the range accepted for $\gamma$ Dor stars which strengthens our confidence in having obtained correct frequency
identifications, particularly for $f_2$ which was unusually low. We also
show
the ratio of the rotational frequency
to the co-rotating frequency which can give an indication to the validity of the mode identification models being applied by FAMIAS. In general
ratios less than 0.5 are considered to lie well within the parameter space which FAMIAS can produce reliable results, however this is dependent on
the value of $m$ considered. \citet{2003MNRAS.343..125T} demonstrated that prograde modes ($m > 0$) are distorted less from increasing Coriolis force,
increasing the limits of reliability. See \citet{2011ApJ...728L..20W} for further discussion.  

Note the results obtained in this section must be regarded as approximate as we have estimated the
stellar mass and radius.

\begin{center}
\begin{table}\caption{Stellar rotational frequencies. Final frequencies, their co-rotational values ($f_{co-rot}$), horizontal-to-vertical
amplitude ratio ($\kappa$, see equation \ref{kvalue}) and ratio of rotational frequency ($f_{rot}$) to each co-rotational frequency.}\label{stepul}
\begin{center}
\begin{tabular}{lllll}
\hline
Frequency & Frequency & $f_{co-rot}$ & $\kappa$ & $f_{rot}$/$f_{co-rot}$ \\
ID & d$^{-1}$ & d$^{-1}$ & & \\
\hline
\\
$f_1$ & 1.3150 & 0.8241 & 40 & 0.6 \\
$f_2$ & 0.2902 & 1.2720 & 17 & 0.4 \\
$f_3$ & 1.4045 & 1.4045 & 14 & 0.3 \\
$f_4$ & 1.8829 & 1.3920 & 14 & 0.4 \\
\hline
\end{tabular}
\end{center}
\end{table}
\end{center}

\section{Discussion}\label{disc}

The dataset in this study provides the largest spectroscopic, and indeed
largest observational, dataset for HD\,135825 to
date. We are now in a position to examine our results in the context of previous work on this star. Multi-periodicy was first confirmed in photometry
by \citet{2002ASPC..256..203E} (Note on page 209 HD\,135825 (HIP\,74825) is incorrectly
written as HD\,135828) but no frequencies were published. One
pulsation period was identified in HIPPARCOS photometry, $0.76053$~d
($1.31487$~d$^{-1}$) \citep{2006AandA...449..281D}. In the same paper $17$
spectra were analysed and one period peak identified at $0.63$~d ($1.59$~d$^{-1}$), but this was treated with caution due to the limited spectra
available at the time.
No mode-identification has previously been published. As mentioned above, the
$v$sin$i$ value of $39.7$ kms$^{-1}$ matches well with previously published
of $38\pm5$\,kms$^{-1}$
\citet{2006AandA...449..281D}

The first frequency extracted in this study matches that found in HIPPARCOS data, but the
$1.59$~d$^{-1}$ frequency was not found.

The high numbers of single-site data can be used to extract multiple frequencies from a g-mode pulsator. It is worth noting that
much of the ambiguity in the path selection and hence frequency identification in the $1^{st}$--$3^{rd}$ moments could be removed using multi-site
data. It is also necessary that any multi-site data is acquired such that it covers the phase range of the frequencies independently. Figure
\ref{radvelfit} shows how multi-site data could distinguish clearly the better frequency fits by providing data points where the fits
differ. Despite the increased difficulty in extracting the pulsations from single-site data, the authors are confident that the frequency
and
mode identifications given in this research are robust, given that similar results are obtained from independent analysis techniques.

We can now begin to remark on the size and quality of spectra required to
undertake geometric pulsation analysis in spectroscopy. The dataset produced
high signal-to-noise frequencies and clear mode identifications using $291$
high-resolution spectra taken over 18 months. It is clear that this is a
demanding standard to study all $\gamma$ Dor stars but the significant
increase in precision and ability to successfully identify multiple frequencies
warrants this approach. It is noted that not all stars will produce similar
results with large datasets. For example, HD\,40745 \citep{2011MNRAS.415.2977M}
has complicated frequencies and modes even using
more than 400 spectra of a comparable quality
to this study. Even more spectra (nearly 700 from multiple sites) were obtained by \citet{2008AandA...489.1213U} on the star HD\,49434 and up to six
frequencies were found in
each of the spectroscopic methods although only some frequencies were found in multiple methods.
 
Single-line analysis of the frequencies
in HD\,135825 produced the same $f_{1}$ in phase for all lines tested. The assumption that all
spectral lines move in phase has been found true for all $\gamma$ Dor stars published to date (see \citet{2011MNRAS.415.2977M},
\citet{NewEntry3} for more examples). This
demonstrates that these stars are suitable for application of the cross-correlation method and this technique is recommended to obtain sufficiently
high 
signal-to-noise to determine multiple frequencies in spectroscopic data. 

As observational asteroseismology of $\gamma$ Dor stars moves beyond the
classification phase into
producing more individual identifications, it seems likely most stars will
require similarly large, or larger datasets as this study. Stars with temperature variations and other periodic structure, such as
tidal effects from binaries, will provide further challenges and also
require massive datasets.

It is clear that high-precision spectroscopic mode identification is dependent on the availability of high-precision spectral abundance
analysis and modelling. The production of the best mode identification is reliant on the availability of T$_{eff}$ and log $g$ measurements and also
benefits from good estimates of stellar masses, radii and inclination. The production of co-rotating frequencies is crucially dependent on precise
values of inclination and radius, which are not well constrained in the current mode-identification method. 

Even small numbers of full mode-identified frequencies can be used to place constraints on stellar models, particularly on the conditions
required to produce such a selection of excited modes. Future work is to take the mode identifications of HD\,135825 from this paper and use
them in complex theoretical models (e.g. \citealt{2003MNRAS.343..125T}) to further constrain stellar parameters and to study the extent of the
core. This could give us important insights
into the evolutionary past and future of the star. Feedback from such models will additionally inform our mode identification methods and lead us ever
closer to understanding these information-rich $\gamma$ Dor stars.

\section{Acknowledgements}
This work was supported by the Marsden Fund.

The authors acknowledge the assistance of staff at
Mt John University of Observatory, a research station of the University of
Canterbury.

We appreciate the time allocated at other facilities for multi-site campaigns, paticularly the Dominion Astrophysical Observatory, McDonald
Observatory and L'Observatoire de Haute Provence.

Gratitude must be extended to the numerous observers who make acquisition of large datasets possible. We thank L.S. Yang Stephenson for observations
taken at the DAO.

This research has made use
of the SIMBAD astronomical database operated at the CDS in
Strasbourg, France.

Mode identification results obtained with the software package FAMIAS developed
in the framework of the FP6 European Coordination action HELAS
(http://www.helas-eu.org/).

P.L.C. acknowledges the hospitality of the MPA from 2011 May- August that enabled him to devote time to various research projects.

We are thankful to our anonymous referee for helpful comments which improved this manuscript.
\label{lastpage}
\bibliography{references}{}
\bibliographystyle{mn2e}
\end{document}